\documentclass[preprint,12pt]{elsarticle}

\usepackage{graphicx}

\newcommand{\be}{\begin{equation}}
\newcommand{\ee}{\end{equation}}
\newcommand{\ba}{\begin{eqnarray}}
\newcommand{\ea}{\end{eqnarray}}

\newcommand{\phys}{\mathrm{phys}}

\journal{Nuclear Physics B}

\begin{document}

\begin{frontmatter}
%%%%%%%% extra for preprint %%%%%%%%%%%%%%%%%%%%%%%%%%%%%%%%%%%%%%%%%%%%%%%%
\begin{titlepage}
\begin{flushright}
LU TP 10-21\\
%arXiv:yymm.nnnn [hep-ph]\\
August 2010
\end{flushright}
\vfill
\begin{center}
{\Large\bf The Massive
       $O(N)$ Non-linear Sigma Model at High Orders}
\\[1cm]
{\bf Johan Bijnens and Lisa Carloni}
\\[1cm]
{Department of Astronomy and Theoretical Physics, Lund University,\\
S\"olvegatan 14A, SE 223-62 Lund, Sweden}
\vfill
{\bf Abstract}\\[1cm]
\end{center}
We extend our earlier work on the massive $O(N)$ nonlinear sigma model
to other observables. We derive expressions
at leading order in the large $N$ expansion at all orders in the
loop expansion
for the decay constant,
vacuum expectation value, meson-meson scattering and the scalar and vector
form factors. This is done using cactus diagram resummation using
a generalized gap equation and other recursion relations.
For general $N$ we derive the expressions for the $n$-th loop order leading
logarithms $\left(M^2/F^2\log(\mu^2/M^2)\right)^n$, up
to five-loops for the decay constant and vacuum expectation value (VEV)
and up to four-loops for meson-meson scattering,  the scalar and vector
form factors. We also quote our earlier result for the mass.
The large $N$ results do not give a good
approximation for the case $N=3$.
We use our results to study the convergence of the perturbative series
and compare with elastic unitarity.
\\[0.5cm]

\end{titlepage}
%%%%%%%%%%%%%%%%%%%%%%%%%%%%%%%%%%%%%%%%%%%%%%%%%%%%%%%%%%%%%%%%%%%%%%%%%%%
\title{The Massive
       $O(N)$ Non-linear Sigma Model at High Orders}
\author{Johan Bijnens and Lisa Carloni}
\address{Department of Astronomy and Theoretical Physics, Lund University,\\
S\"olvegatan 14A, SE 223-62 Lund, Sweden}

\begin{abstract}
We extend our earlier work on the massive $O(N)$ nonlinear sigma model
to other observables. We derive expressions
at leading order in the large $N$ expansion at all orders in the
loop expansion
for the decay constant,
vacuum expectation value, meson-meson scattering and the scalar and vector
form factors. This is done using cactus diagram resummation using
a generalized gap equation and other recursion relations.
For general $N$ we derive the expressions for the $n$-th loop order leading
logarithms $\left(M^2/F^2\log(\mu^2/M^2)\right)^n$, up
to five-loops for the decay constant and vacuum expectation value (VEV)
and up to four-loops for meson-meson scattering,  the scalar and vector
form factors. We also quote our earlier result for the mass.
The large $N$ results do not give a good
approximation for the case $N=3$.
We use our results to study the convergence of the perturbative series
and compare with elastic unitarity.
\end{abstract}

% {\bf PACS:} 11.10.Hi, Renormalization group evolution of parameters, 
% 11.15.Pg, Expansions for large numbers of components (e.g., $1/N$ expansions) 
% 11.30.Qc, Spontaneous and radiative symmetry breaking 
% 12.39.Fe  Chiral Lagrangians 


\end{frontmatter}

%%%%%%%%%%%%%%%%%%%%%%%%%%%%%%%%%%%

\section{Introduction}
\label{intro}

In a renormalizable field theory, given enough time and computer power,
one may calculate an observable up to any order $n$ in the perturbative
expansion,  study the convergence of the series and make a precision
comparison with experimental results. In the absence of complete higher order
calculations, a first estimate of the convergence of the perturbative series
may  
come from the so-called leading logarithms (LL). These are terms of the form
$\alpha^n\log^n\mu^2$, with $\mu$ the renormalization scale, which appear in
 the $n$-th order corrections of any observable´s expression upon
renormalization. 
The coefficients of these LL may be calculated using renormalization group
methods. 

In a non-renormalizable field theory the situation is more convoluted,
since new terms appear in the Lagrangian at each order in the expansion.
However, as Li and Pagels pointed out \cite{LiPagels} to one-loop,
the  $n$-loop-order contributions will still 
contain a $c_n\log^n(\mu^2/M^2)$ term. 
Consider for example the expansion of the pion mass in the quark masses
in Chiral 
Perturbation Theory \cite{GL1}:
\be
\label{eq:mass} 
M^2_{\pi}=M^2\left[
 1+\frac{M^2}{(4 \pi F)^2}\left(-\frac{1}{2}\log{\frac{M^2}{\mu^2}}
 +\ell^r_3(\mu)\right)+\cdots\right]
\ee
where $\ell^r_3$ is a renormalized second order Lagrangian coupling
and $M$ and $F$ are the lowest order Lagrangian parameters.
Depending upon the $\mu^2/M^2$ ratio, where the parameter $M^2$ 
is the lowest-order mass and $\mu$ is the renormalization scale,
the term with the logarithm can be the largest part of the correction.
In general the LL, now depending on a typical scale $\mathcal{M}$
of the process one looks at, 
may turn out to be a substantial fraction of the $n$-th-order correction.
For many observables indeed the LL are the 
main contribution due to the enhancement of $\log(\mu^2/M^2)$ compared
to the other contributions at the same order. 
This is true in particular for the $\pi\pi$ $S$-wave scattering length
$a^0_0$ \cite{GL1,GL0} in Chiral Perturbation Theory 
(ChPT).  

Many corrections of this type at one-loop have been calculated long ago, 
see the review \cite{Pagels} and references 
therein, mostly in the framework of Current Algebra,
and in ChPT \cite{GL1,Weinberg1,GL2}.

Weinberg pointed out in 1979 \cite{Weinberg1} that, because of
renormalization group 
equations (RGE),  the two-loop LL coefficient appearing in $\pi \pi$
scattering amplitude 
could be calculated using simple one-loop diagrams. 
This method was later used  in \cite{Colangelopipi} for two loop 
LL corrections to scattering lengths and slopes in $\pi \pi$ scattering
and in \cite{BCE0} for the general three-flavour meson sector.
Nowadays the extension to the full two-loop expressions 
for mesons is known for most observables \cite{Bijnens1}.

Weinberg's renormalization argument was extended  
by B\"uchler and Colangelo~\cite{BC} 
to all orders and to a generic non-renormalizable theory.
They showed that the leading logarithms at any loop-order
can be calculated using one-loop diagrams. 
They also showed that the coefficient of the leading logarithm
only depends upon the constants appearing in the lowest-order Lagrangian.
In principle this coefficient at $n$-loop-order could have depended on
all of the coupling constants in the Lagrangians $\mathcal{L}_m$
with $m<n$.   
 
However, the problem remains that as $n$ grows,  the number of terms 
and counterterms in the Lagrangian grows very rapidly, and the
renormalization group 
equations (RGE) become more involved. This renders the calculation
of LL beyond 
the first few orders a Herculean task\footnote{See for example the two-loop
leading logarithm in the non leptonic sector \cite{Buchler2}.}.

The alternative to performing these long calculations is to extract
the LL series from 
a renormalizable theory in hopes that it will  reproduce the LLs of
the non-renormalizable theory. 
The authors of \cite{BF2} applied this approach to the renormalizable
linear sigma model 
and were able to resum the entire LL series by exploiting recursion relations. 
They however found recursive relations were not possible in the 
non-renormalizable non-linear sigma case \cite{BF2}.

In the massless case, a solution to managing the terms in the Lagrangian
was found 
since the number of meson legs on the vertices one needs to consider remains 
limited \cite{Kivel1}. This was used for meson-meson scattering
\cite{Kivel1} and the scalar and vector form factor \cite{Kivel2} in the
massless $O(N)$ model. This method works to arbitrarily high order
and agrees with the known large $N$ \cite{ColemanlargeN}
results to all orders \cite{Kivel1,Kivel2}. 
In the massless case, one may also use kinematic methods to extract the
nonanalytic 
dependence on kinematic quantities. These have for instance been used to
derive the 
form factors \cite{BF1} up to five loop-order and to arbitrarily high order for 
both the form factors and the meson-meson scattering amplitude \cite{Kivel3}. 
These methods essentially solve the leading logarithm
problem in the massless case for the most useful observables.

In the massive case (which includes ChPT), however, tadpole diagrams no 
longer vanish and one needs to consider terms with
an increasing numbers of meson legs. E.g. for the meson mass
one needs to calculate one-loop diagrams with $2n$ meson legs in order
to get the  $n$-loop-order LL. In our earlier work \cite{paper1} we showed 
that one does not need to explicitly construct the higher order Lagrangians in 
a symmetric form, nor does one need a minimal Lagrangian at each order. 
The LL series only requires order by order a complete enough Lagrangian  
to describe the observable at hand \cite{paper1}.  
This means that one may let the algorithm itself generate all the necessary
terms in the higher order Lagrangians.

 We applied this method to obtain the leading logarithm to five-loop-order for
the meson mass in the nonlinear $O(N)$-model. One set of results of this
paper is to extend the calculation to the decay constant and vacuum expectation
value also to five-loop-order.
A similar amount of work allows to obtain the leading logarithms
for meson-meson scattering and the scalar and vector form factors to
four-loops and we present results for this as well. For $N=3$ the massive
non linear $O(N)$ model is equivalent to two-flavour mesonic ChPT Lagrangian
at lowest order in the sense that $O(4)/O(3)$
is isomorphic to $SU(2)_L\times SU(2)_R/SU(2)_V$.
We have thus obtained the leading logarithms also for this
physically interesting case to rather high loop-order.

In \cite{paper1} we also extended the large $N$ limit
to the massive case and applied it for the meson mass.
In this paper we extend those methods to many more observables and we
find for all the cases considered simple expressions in terms of the
physical quantities. However, as already observed in the
massless case \cite{Kivel1,Kivel2} and for the mass \cite{paper1}, we again
find that the large $N$ result does not give a good approximation to the
coefficient of the leading logarithm for general $N$ and in particular
not for $N=3$.

We briefly summarize the methods
for calculating leading logarithms of  \cite{BC,paper1}
in Sect.~\ref{BCsummary} and introduce the massive $O(N)$ nonlinear sigma model
including external fields in Sect.~\ref{sigmamodel}. 
Here we also define all the physical quantities we calculate.
The next section
discusses the large $N$-limit. We briefly recall the results of \cite{paper1}
and extend the method to the other observables. The leading logarithms for
general $N$ are discussed in Sect.~\ref{LLseries}. There we also discuss the
convergence of the various observables and compare some ways of
expanding. The main results and conclusions are summarized in the last section.

One note about the cross-checks on our results. All calculations were
performed using four different parametrizations for the fields. This means
that for every parametrization the form of the Lagrangian and the couplings
are different. These four different Lagrangians were fed into the same form
FORM code \cite{FORM}. The fact that the output for all observables came out
the same regardless of the parametrization is a very good sanity check.
 
\section{Counter terms and Leading Logarithms}
\label{BCsummary}

In this section we present the results of \cite{BC} and \cite{paper1}. We show 
how to calculate the LL coefficients and their connection to the counter terms 
used to renormalize the observable.

In order to calculate the matrix elements for the observables, consider
the generating functional 
\be
W[j]=e^{iZ[j]}/\hbar=\int \mathcal{D}\phi_ie^{iS[\phi,j]/\hbar}.
\ee
where $j$ is the classical source which allows one to extract all
of the Green functions. The action $S$ may be expanded around the classical 
action, $S=\sum^{\infty}_{n=0}\hbar^nS^{(n)}$. 
In practice one expands the Lagrangian into a
sum of growing $\hbar$ order (bare) Lagrangians 
\be
\mathcal{L}=\sum^{\infty}_{n=0}\hbar^n\mathcal{L}^{(n)}.
\ee
The crucial difference between renormalizable and
non-renormalizable theories is the 
number of terms appearing in
each $\mathcal{L}^{(n)}=\sum^{N_n}_i c^{(n)}_iO^{(n)}_i $.
In the former case the terms in all $\mathcal{L}^{(n)}$
are of the same form as in  $\mathcal{L}^{(0)}$.
In the later case, new terms appear at each order $n$. 
When calculating matrix elements beyond tree level,
the loop corrections lead to divergences, 
which must be re-absorbed into a redefinition of the coupling constants.
While for a renormalizable theory it suffices to reabsorb
them into lowest order Lagrangian since they are all of the same form,
in a non-renormalizable
theory a divergence is absorbed by the higher order coupling constants.
E.g. the divergence that appeared in the calculation
of  (\ref{eq:mass}) was absorbed into $\ell_3$. 
Alternatively, one may say that the renormalization consists in adding to the
Lagrangian order by order operators $\mathcal{O}^{(n)}_i$ with diverging
coefficients to cancel divergences,
i.e. counterterms  $c^{(n)}_{ik}\mathcal{O}^{(n)}_i/\epsilon^{k}$. 
\be
\mathcal{L}^{(n)}=
\frac{1}{\mu^{\epsilon n}}
\left[ \mathcal{L}^{(n),\mathrm{ren}} +\mathcal{L}^{(n), \mathrm{div}}\right]
=\frac{1}{\mu^{\epsilon n}}
\left[c^{(n)}_{i0}(\mu)\mathcal{O}^{(n)}_i 
+\sum^n_{k=1}\frac{c^{(n)}_{ik}\mathcal{O}^{(n)}_i}{\epsilon^{k}}\right]\,.
\ee
where we have assumed one works in dimensional regularization\footnote{The
pre-factor $\frac{1}{\mu^{\epsilon n}}$ ensures that all Lagrangians have
the same dimension $d$.},
in which divergences appear as poles $1/\epsilon^{k}$, $\epsilon=4-d$. 
We have shown in \cite{paper1} that set operators $\mathcal{O}^{(n)}_i$ need
not be minimal or even complete for our purposes.

Consider an observable and the one particle irreducible (1PI) diagrams
that may contribute at 
order $n$. That one only needs to consider 1PI diagrams was proven
in \cite{BC}. Let 
$L^n_\ell$ be the  $n$-th order contribution  from $\ell$ loops. $L^n_\ell$
consists of a finite part 
and (possibly) several different poles 
\be
L^n_\ell= L^n_{\ell 0}+\sum^l_{k=1}  \frac{L^n_{\ell k}}{\epsilon^k}.
\ee
In $L^n_{\ell k}/\epsilon^k$ we include only those divergences coming from
the loop integration 
and not those coming from the diverging $c^{(m)}_{ik}\mathcal{O}_i$ vertices
in the loops. Each 
loop may contain different $c^{(m)}_{ik}\mathcal{O}_i$ vertices. We indicate
with 
$\{ c\}^n_\ell$ the product
$c^{(m_1)}_{i_1k_1} c^{(m_2)}_{i_2k_2}\cdots c^{(m_r)}_{i_rk_r}$  giving 
an $n$-th order, $\ell$-loop contribution. 
 
The recursive equations follow from the requirement that the divergences
must cancel.
The contribution at order $\hbar$
may be written as
\be
\label{one-loop}
\frac{1}{\epsilon}\left[ \frac{1}{\mu^\epsilon}L^1_{00}\left(\{ c\}^1_1\right)
+L^1_{11}\right]
+\frac{1}{\mu^\epsilon}L^1_{00}\left(\{ c\}^1_0\right)+L^1_{10}.
\ee
where $L^1_{11}=L^1_{11}\left(\{ c\}^0_0\right)$. Expanding 
$\mu^{-\epsilon}=1-\epsilon \log{\mu}+\cdots$, one finds
that to cancel the $1/\epsilon$ we need
\be
\label{betaoneloop}
L^1_{00}\left(\{ c\}^1_1\right)=-L^1_{11}\,.
\ee
This determines all the needed $c^{(1)}_{i1}$ in terms of the lowest order
parameters. From (\ref{one-loop}) we also get the explicit dependence on $\mu$
\be
-L^1_{00}\left(\{ c\}^1_1\right) \log{\mu} = L^1_{11}\log{\mu}
\ee
where the equality follows from (\ref{betaoneloop}).
To summarize, the counterterm $c^{(1)}_{11}$ is adjusted so that it
cancels the divergence coming from 
the loop, $L^1_{11}$, this in turn determines the
$\mu$ dependence and hence the coefficient of the LL. 
At second order
the cancellation of the $1/\epsilon^2$  and $\log(\mu)/\epsilon$
pieces allow to obtain the leading divergence $c^{(2)}_{i2}$
from the one-loop part $L^2_{11}(\{c\}^1_1)$
and expanding $\mu^\epsilon$ one finds that 
the coefficient of the LL $\log^2(\mu)$, is $L^2_{22}$. See \cite{paper1}
for a more detailed discussion.

These results may be generalized. At order $n$ one may write
\be
\frac{1}{\epsilon^n}
\left[\frac{1}{\mu^{n\epsilon}}L^n_{00}\left(\{ c\}^n_n \right)+
\frac{1}{\mu^{(n-1)\epsilon}}L^n_{11}\left(\{ c\}^{n-1}_{n-1}\right)+\cdots + 
\frac{1}{\mu^{\epsilon}}L^n_{n-1,n-1}\left(\{ c\}^1_1 \right)+L^n_{nn}\right].
\ee
Requiring that the coefficients of $1/\epsilon^n, \log{\mu}/\epsilon^{n-1},
\log^2\mu/\epsilon^{n-2},\ldots$ cancel 
leads to a set of $n$ equations,
the solution of which is given by \cite{paper1}
\be
L^n_{n-i,n-i}\left( \{ c\}^i_i \right)=(-1)^i
\left(\begin{array}{c}n\\i\end{array}\right)L^n_{nn}.
\ee
In particular, 
\be
L^n_{11}\left( \{ c\}^{n-1}_{n-1} \right)=(-1)^{n-1}nL^n_{nn}
\label{eq:recursive}
\ee 
and
\be
L^n_{11}\left( \{ c\}^{n-1}_{n-1} \right)= (-n)L^n_{00}
\label{hbarnsol}
\ee 

The coefficient of the leading logarithm is given by
 \be
L^n_{nn} \left(\log{\mu}\right)^n.
\ee
Eq. (\ref{eq:recursive}) is solved recursively. 
First one calculates the one loop counterterm.  
With this one, using  (\ref{eq:recursive}), 
one may calculate $L^2_{22}$,
the coefficient of the second order LL.
This again fixes the $c^{(2)}_{i2}$ counterterm, which can be inserted
back into Eq. (\ref{eq:recursive}), and so on. One only needs to insure
that all the $c^{(n-1)}_{i,n-1}$ for all the $\mathcal{O}^{(n-1)}_i$ 
appearing in the calculation are determined.
 
\section{Massive nonlinear $O(N+1)/O(N)$ sigma model}
\label{sigmamodel}

The $O(N+1)/O(N)$ nonlinear sigma model, including external sources\footnote{If one wishes to study a given current $J^\mu$ one adds an extra classical source field $v_\mu$ to the generating functional which couples to that current. Thus the generating functional becomes
$$
W[j, v_\mu]=\int {\mathcal{D}\phi e^{i\int{(\mathcal{L}-j\phi-v_{\mu}}J^\mu})/\hbar}.
$$
The matrix elements involving $J^\mu$ can be obtained by functional derivation with respect to $v_\mu$ 
$$
J^\mu(x)=\left.\frac{\delta \log{W}}{\delta v_\mu(x)}\right|_{v_\mu=0}.
$$
}, is described by the Lagrangian
\be
\label{sigmalag}
\mathcal{L}_{n\sigma} = \frac{F^2}{2}
D_\mu \Phi^T D^\mu \Phi+F^2 \chi^T \Phi\,,
\ee
where $\Phi$ is a real $N+1$ vector,
$\Phi^T =\left(\Phi^0~\Phi^1~\ldots~\Phi^N\right)$, 
which transforms as the fundamental
representation of $O(N+1)$ and satisfies the constraint $\Phi^T\Phi=1$.
The covariant derivative is given by
\ba
D_\mu \Phi^0 &=&\partial_\mu \Phi^0 + a^a_\mu \Phi^a\,,
\nonumber\\
D_\mu \Phi^a &=&\partial_\mu\Phi^a +v^{ab}\Phi^b-a^a_\mu\Phi^0\,.
\ea
The vector sources satisfy $v^{ab}_\mu=-v^{ba}_\mu$ and correspond to the unbroken
group generators while the axial $a^a_\mu$ sources correspond to the broken ones.
Indices of the type $a,b,\ldots$ run over $1,\ldots,N$ in the remainder.
The mass term 
$\chi^T \Phi$
contains the scalar, $s^0$, and pseudo-scalar, $p^a$ external fields as well
as the explicit symmetry breaking term $M^2$.
\be
\chi^T =\left(~(2Bs^0+M^2)~~p^1~\ldots~p^N\right)\,.
\ee
The term proportional to $M^2$
breaks the $O(N+1)$ symmetry explicitly to the $O(N)$,
whereas the  vacuum condensate 
\be
\label{vacuum}
\langle \Phi^T \rangle = \left(1~0~\ldots~0\right)\,,
\ee
breaks it spontaneously to the same $O(N)$.

This particular model corresponds to lowest 
order two-flavour ChPT for $N=3$ \cite{GL1,Weinberglag}.
It is also used as a model for a strongly interacting Higgs sectors
in several beyond Standard models,
see e.g. \cite{custodial1,custodial2}.

The terminology for the external sources or fields is taken from
two-flavour ChPT. The vector currents for $N=3$ are given by
$v^{ab} = \varepsilon^{cab} v^c$ with $\varepsilon^{cba}$ 
the Levi-Civita tensor. The electromagnetic current
at lowest order is associated to $v^3$. The external scalar source $s^0$ 
is instead associated to the  QCD current $-\overline u u-\overline d d$ as in \cite{GL1}.

We write $\Phi$ in terms of a real $N$-component\footnote{We refer to these
as a flavour components.} vector $\phi$, which transforms linearly under
the unbroken part of the symmetry group, $O(N)$. 
In the calculations, for simplicity, we will refer
to one particular parametrization, called $\Phi_1$ below.
We have, however, made use of four different parametrizations in order
to check the validity of our results. These are
$$
\begin{array}{ccc}
%\label{parGL}
\Phi_1 =
\left(\begin{array}{c}\sqrt{1-\frac{\phi^T\phi}{F^2}}\\ \frac{\phi}{F}\end{array}\right)\,
&&
%\label{parWE}
\Phi_2 = \frac{1}{\sqrt{1+\frac{\phi^T\phi}{F^2}}}
\left(\begin{array}{c}1\\\frac{\phi}{F}\end{array}\right)\,\\
%\label{par3}
\Phi_3 = \left(\begin{array}{c}1-\frac{1}{2}\frac{\phi^T\phi}{F^2}
\\[2mm]
\sqrt{1-\frac{1}{4}\frac{\phi^T\phi}{F^2}}\frac{\phi}{F}\end{array}\right)\,
&&
%\label{par4}
\Phi_4 = \left(\begin{array}{c}\cos\sqrt{\frac{\phi^T\phi}{F^2}}
\\[2mm]
\sin\sqrt{\frac{\phi^T\phi}{F^2}}\,\frac{\phi}{\sqrt{\phi^T\phi}}\end{array}\right)
\,.
\end{array}
$$

$\Phi_1$ is the parametrization used in \cite{GL1}, 
$\Phi_2$ a simple variation,
$\Phi_3$ is such that the explicit symmetry breaking term in (\ref{sigmalag})
only gives a mass term to the $\phi$ field but no vertices.
$\Phi_4$ is the parametrization one ends up with if using the general
prescription of \cite{CWZ}.

The physical mass of the meson (squared), $M^2_\phys$, we already calculated 
in \cite{paper1}.
The meson decay constant, $F_\phys$, is defined by the matrix element of
the axial current $j_{a,\mu}^b$
\be
\label{defF}
\langle 0| j^b_{a,\mu}| \phi^c(\mathbf{p})\rangle = i F_\phys
\mathbf{p}_\mu\delta^{bc}\,. 
\ee
The lowest order is $F_\phys=F$.

The vacuum expectation value (VEV) is defined by
\be
\label{defV}
V_\phys = \langle -j^0_{s^0}\rangle \stackrel{N=3}{=}
 \langle \overline u u +\overline d d\rangle\,.
\ee
In the second equation we have written out its meaning in terms of quarks.
Its lowest order value is $V_\phys\equiv V_0 = -2BF^2$.

The scalar form factor is defined as the matrix element of the scalar current
between two mesons
\be
\label{defFS}
\langle\phi^a(\mathbf{p}_f)|-j^0_{s^0}|\phi^a(\mathbf{p}_i)\rangle
= F_S[ (\mathbf{p}_f-\mathbf{p}_i)^2]\,.
\ee
The value at zero momentum transfer can be derived from the meson mass 
via the Feynman-Hellmann
theorem as
\be
\label{FeynmanHellmann}
F_S(0) = 2B\frac{\partial M^2_\phys}{\partial M^2}\,.
\ee
The vector form factor is defined similarly as
\be
\label{defFV}
\langle\phi^a(\mathbf{p}_f)|
  j^{cd}_{V,\mu}-j^{dc}_{V,\mu}|\phi^{b}(\mathbf{p}_i)\rangle
= \left(\delta^{ac}\delta^{db}-\delta^{ad}\delta^{bc}\right)
    i(\mathbf{p}_f+\mathbf{p}_i)^\mu 
  F_V\left[(\mathbf{p}_f-\mathbf{p}_i)^2\right]\,,
\ee
where we have exploited the antisymmetry of $v^{ab}_\mu $. 
The vector currents $j^c_V$ for the $N=3$ are given
by $j^{ab}=j^c_V\varepsilon^{cab}$ with 
$\varepsilon^{cab}$ the Levi-Civita tensor.
The electromagnetic current in this case to 
the lowest order is given by 
$j^3_{V,\mu}=(\bar{u}\gamma_\mu u -\bar{d}\gamma_\mu d)/2$.
The singlet part does not appear to lowest order.
 The value of the vector 
form factor at $(\mathbf{p}_f-\mathbf{p}_i)^2=0$ is always 
1 because of the conserved $O(N)$ symmetry. 

In addition to the form factors we also define the radii and curvatures
with $t=(\mathbf{p}_f-\mathbf{p}_i)^2$ and expanding around $t=0$:
\ba
\label{defradii}
F_S(t) &=& F_S(0)\left(1+\frac{1}{6}\langle r^2\rangle_S t
 + c_S t^2+\cdots\right)\,.
\nonumber\\
F_V(t) &=& 1+\frac{1}{6}\langle r^2\rangle
_V t + c_V t^2+\cdots\,.
\ea

The final process we discuss is meson-meson scattering. The general amplitude
is
\ba
\label{defAstu}
\langle\phi^a(\mathbf{p}_a)\phi^b(\mathbf{p}_b)|
    \phi^c(\mathbf{p}_c)\phi^d(\mathbf{p}_d)\rangle
&=& \delta^{ab}\delta^{cd} A(s,t,u)+\delta^{ac}\delta^{bd} A(t,u,s)
\nonumber\\&&
  +\delta^{ad}\delta^{bc} A(u,s,t)\,,
\ea
with
\be
s = (\mathbf{p}_a+\mathbf{p}_b)^2,\qquad t
 =(\mathbf{p}_a-\mathbf{p}_c)^2\qquad u = (\mathbf{p}_a-\mathbf{p}_d)^2\,,
\ee
satisfying $s+t+u=4 M^2_\phys$. $A(s,t,u)$ is symmetric in $t$ and $u$.
The proof of (\ref{defAstu}) for $N=3$ is done using crossing and
isospin symmetry
\cite{Weinbergpipi}, but may be generalized to the $O(N)$ symmetry case.
There are three channels, the singlet, antisymmetric
and symmetric combination, named $I=0,1$ and 2 (after isospin conservation).
The amplitude in these three channels is given by
\ba
\label{defTI}
T^0(s,t)&=&N A(s,t,u)+A(t,u,s)+A(u,s,t)\nonumber\\
T^1(s,t)&=&A(t,u,s)-A(u,s,t)\nonumber\\
T^2(s,t)&=&A(t,u,s)+A(u,s,t)\,.
\ea
These amplitudes are projected onto the partial waves
\be
\label{defTIl}
T^I_\ell
= \frac{1}{64 \pi}\int^1_{-1}d(\cos \theta) P_{\ell}(\cos \theta) T^I(s,t)\,,
\ee
with $\theta$ the scattering angle and $P_\ell$ the Legendre polynomials.
Near threshold the partial waves can be expanded in terms of scattering lengths
$a^I_{\ell}$ and slopes $b^I_{\ell}$.
\be
\label{defscatt}
\Re \left[T^I_{\ell}\right]= q^{2\ell}\left(a^I_{\ell}+b^I_{\ell}q^2
        +\cdots\right),
\ee
where $q^2\equiv\frac{1}{4}\left(s-4M^2_\phys\right)$.
The scattering lengths and slopes are normally given in units of powers
of $M_\phys$.

For all the quantities defined here, the $N=3$ results with a 
complete $O(p^6)$ Lagrangian 
are known up to two loops, thus including the leading logarithms.
This is an additional check 
on our calculation.

The results can be expressed in terms of the lowest order parameters,
expanding in powers of 
\be
\label{defL}
L = \frac{M^2}{16\pi^2 F^2}\log\frac{\mu^2}{M^2}\,,
\ee
or in terms of the physical mass and physical decay constant using 
\be
\label{defLphys}
L_\phys = \frac{M^2_\phys}{16\pi^2 F^2_\phys}\log\frac{\mu^2}{M^2_\phys}\,.
\ee
In both cases we chose the mass scale in the logarithm to be the
corresponding mass.
In \cite{paper1} we also used an expansion in
\be
\label{defLMphys}
L_{M_\phys} = \frac{M^2_\phys}{16\pi^2 F^2}\log\frac{\mu^2}{M^2_\phys}\,.
\ee
We use this hybrid form in this paper only for one figure.

\section{Large $N$ limit}
\label{largeN}

When we consider the limit of large $N$ we have to decide how the constants
in the Lagrangian (\ref{sigmalag}) depend on $N$. The first term can be made
linear (extensive) in $N$ by assuming $F^2\propto N$ since $\Phi^T\Phi=1$.
The second term is then also linear in $N$ if we set $M^2$ and $B$ to be
independent of $N$.

In the linear $O(N)$ model it is well known \cite{ColemanlargeN,custodial1}
that the leading contribution in $N$ comes from diagrams that contain only
non-overlapping loops and in which each factor $1/N$ coming from a new
interaction is canceled by the factor $N$ introduced by summing over
the $N$ internal mesons in each bubble.
In \cite{paper1} we showed how this is also true in general for the non linear
case which has vertices with any number of meson fields.
The proof in \cite{paper1} remains valid when vertices with external
fields are included. The vertices with more mesons legs are suppressed by powers
of $1/F$ as was the case for the purely mesonic vertices. The suppression factor
compared to the lowest order of $1/F^{2L}$ with $L$ the number of loops
remains thus valid as well.

\begin{figure}
\begin{center}
\includegraphics[width=0.3\textwidth]{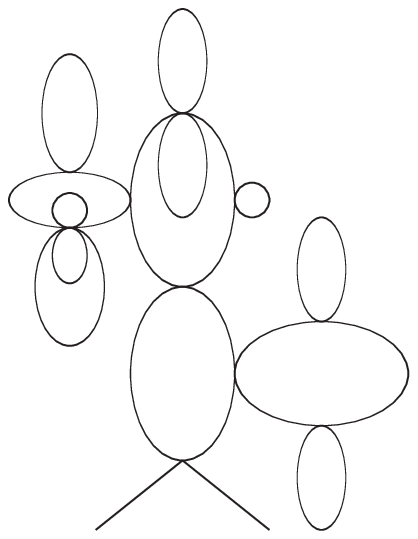}
\end{center}
\caption{\label{figcactus} A typical diagram which contributes at leading
order in $N$. Note that vertices can have many different loops attached
since the Lagrangians contain vertices with any number of fields.
The flavour-loops coincide with momentum loops, i.e. the visible bubbles.}
\end{figure}

In Fig.~\ref{figcactus} we show such a typical diagram contributing to the meson
self energy. All of these diagrams can be resummed by exploiting
recursive methods.

\begin{figure}
\begin{center}
\includegraphics[width=\textwidth]{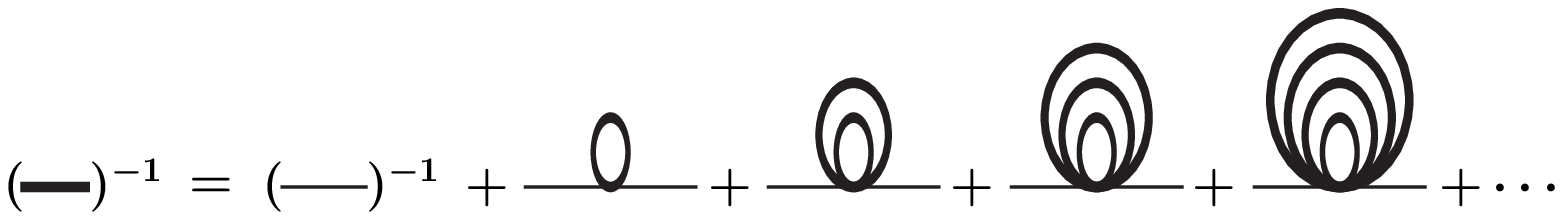}
\end{center}
\caption{\label{figgap} The graphical representation of the equation
that generates all the cactus diagrams for the propagator.
A thick line indicates the full propagator, a thin line the lowest order one.}
\end{figure}
Consider for instance the inverse of the full propagator, it is given by the
inverse of the lowest order propagator and the sum of all the 1PI diagrams
with two external legs. By starting out with the lowest order propagator
on the right hand side (rhs) and then reinserting the solution recursively we
generate all cactus diagrams. In \cite{paper1} we used this method to show
that the full propagator in parametrization 1 in the large $N$ limit is
\be
i\Delta_{\rm{full}}(\mathbf{q}^2) = \frac{i}{\mathbf{q}^2-M^2_\phys}\,.
\ee
Note that, as shown in \cite{paper1} and below,
in this parametrization in this limit there is no wavefunction
renormalization. Let us note once more that in the following we derive the 
results in the first parametrization but the others can also be used and give
the same results.

This method is similar to the gap equation used in e.g. NJL
 models \cite{ENJLreview} but we have here an infinite number of terms
on the right hand side. 
Similar resummations may be used for the other observables as shown below.

For completeness we quote the result for the physical mass \cite{paper1},
it is the solution of
\be
\label{masslargeN}
M^2 = M^2_\phys\sqrt{1+\frac{N}{F^2}\overline A(M^2_\phys)}\,,
\ee
with
\be
\overline A(M^2) = \frac{M^2}{16\pi^2 }\log\frac{\mu^2}{M^2}\,.
\ee

Note that in non renormalizable field theories the large $N$ limit also depends
upon how $\mathcal{L}^{(n)}$ depend on $N$. The result (\ref{masslargeN}) and 
those derived below  assume that the finite part of any higher order 
coefficient vanishes, $c^{(n>0)}_{i0}=0$, see \cite{WeinberglargeN}
for a discussion. 
The formulas still give the leading logarithms even for non zero higher 
order coefficients as long as the Lagrangian remains at most linear in $N$.

\subsection{Decay constant}

\begin{figure}
\begin{center}
\includegraphics[width=\textwidth]{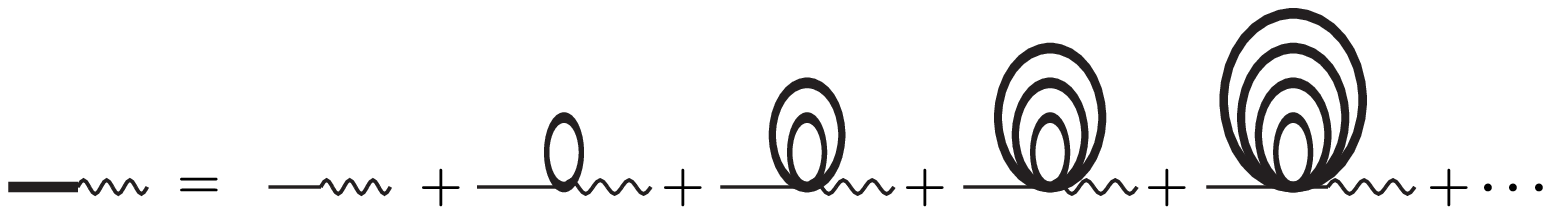}
\end{center}
\caption{ The graphical representation of the equation
that generates all the one-particle-irreducible 
cactus diagrams for the decay constant.
A thick line indicates the full propagator. The photon line indicates
the insertion of the axial current $a^a_\mu$.}
\label{fig:decay_gap}
\end{figure}
{}From the Lagrangian (\ref{sigmalag}) we can extract the vertices
involving the axial current $j^\mu_a$. These are given by
$F^2 (\partial^{\mu} \Phi^0 a^a_{\mu} \Phi^a 
 - \partial^{\mu}\Phi^a a^a_{\mu}\Phi^0)$.
In the first representation $\Phi_1$, this leads to the couplings 
\be
\label{lagrangianaxial}
a^a_{\mu}F\left[
\phi^a\partial_{\mu}\sqrt {1-\frac{\phi^T \phi}{F^2}}
        -\partial_{\mu} \phi^a\sqrt{1-\frac{\phi^T \phi}{F^2}}
\right].
\ee
All 1PI cactus diagrams contributing to the
matrix element in (\ref{defF}) can be generated by the
diagrams with full propagators shown in Fig.~\ref{fig:decay_gap}.
Each of the tadpole loops in Fig.~\ref{fig:decay_gap} must  contribute a
factor of
$N$, to be leading in $N$, thus  it must be generated by the contraction of a
$\phi^T\phi=\phi^b\phi^b$ pair, with a sum over the flavour index $b$. 
All other contractions only give subleading powers in $N$. 

The first term in (\ref{lagrangianaxial}) then gives only terms
with at least one loop integral that vanishes since it is odd in momentum.

We are thus left with the term 
\be
-F \partial_{\mu} \phi^a \mathcal{A}^{\mu}\sqrt{1-\frac{\phi^T \phi}{F^2}}
\simeq -F \partial_{\mu} \phi^a \mathcal{A}^{\mu}
\left[ 1- \frac{\phi^b \phi^b}{2F^2}+...\right]\,.
\ee
When the $\phi^b$ are contracted, they give $i^2$  times the loop integral
$A(M^2_\phys)= (1/i)\int \frac{d^dp}{(2 \pi)^d}/(p^2-M^2_\phys)$.
Note that the mass in this expression is the physical mass and that the 
propagators in Fig. \ref{fig:decay_gap} in the loops are the full propagators. 

We now show that the  wave function renormalization vanishes
in para\-metrization 1.
The part of the Lagrangian that can produce momentum dependence
in the full propagator is given by
\be
\mathcal{L}_{\rm{kin}}=\frac{F^2}{2}\partial_{\mu}\Phi^T \partial^{\mu}\Phi
% \nonumber\\
= \frac{F^2}{2}\partial_{\mu}\sqrt{1-\frac{\phi^T \phi}{F^2}}
        \partial_{\mu}\sqrt{1-\frac{\phi^T \phi}{F^2}}
  +\frac{1}{2}\partial_{\mu}\phi^T \partial^{\mu}\phi. 
\ee
The corrections to the canonical kinetic term come
from the first term, which can be rewritten as
\ba
\mathcal{L}^{\rm{corr}}_{\rm{kin}}
&=&\frac{1}{2F^2}
\frac{(\phi^a \partial_{\mu} \phi^a)(\phi^b \partial^{\mu} \phi^b)}
 {1-\frac{\phi^T \phi}{F^2}}
\nonumber\\
&=&\frac{1}{2F^2}(\phi^a \partial_{\mu} \phi^a)(\phi^b \partial^{\mu} \phi^b)
(1+\frac{\phi^c \phi^c}{F^2}+\cdots)\,.
\ea
In order to have a non-zero loop diagram the derivatives must either
both act on internal legs or both on external legs. 
Either way, since the contracted legs must have the same flavour, $a\equiv b$,
so there can be no sum over the flavour index and thus no leading in $N$
correction. This means that in the large $N$ approximation
 one has $Z=1$ in this parametrization (this will not be true in general).

Putting the terms together, we find the physical decay constant to be
related to the low energy constants $F$ and $M^2$ by the simple relation
\be
\label{decaylargeN}
F_\phys=F\sqrt{1+\frac{N}{F^2}A(M^2_\phys)}.
\ee
To get the LL series one should expand the square root,
replace $A(M^2_\phys)$ by 
$\overline{A}=\frac{M^2_\phys}{16 \pi^2}\log{\mu^2/M^2_\phys}$
and express the $M^2_\phys$ in terms of
$L=\frac{M^2}{16 \pi^2 F^2}\log\frac{\mu^2}{M^2}$.

To express this result in terms of the physical $F_\phys$ and $M^2_\phys$
instead, 
 we can square (\ref{decaylargeN}) to obtain
\be
1+\frac{N}{F^2}\overline A(M^2_\phys) = \frac{1}{\displaystyle
1-\frac{N}{F^2_\phys}\overline A(M^2_\phys)}, 
\ee
which allows us to rewrite (\ref{masslargeN}) and (\ref{decaylargeN}) as
\ba
\label{masslargeN2}
\label{decaylargeN2}
M^2_\phys &=& M^2\sqrt{1-\frac{N}{F^2_\phys}\overline A(M^2_\phys)}\,,
\nonumber\\
F_\phys &=& \frac{F}{\sqrt{1-\frac{N}{F^2_\phys}\overline A(M^2_\phys)}}\,.
\ea

\subsection{Vacuum Expectation Value}

In a similar fashion one can calculate the leading $N$ vacuum expectation 
value series. Consider the second term in (\ref{sigmalag})
\ba
F^2 \chi^T \Phi.
\ea 
where $\chi^T=2 B_0(s^0~~0~\cdots~0)$. 
In the first representation this becomes
\ba
2 B F^2 \sqrt{1-\frac{\phi^T \phi}{F^2}}
= 2 B F^2\left( 1-\frac{1}{2}\frac{\phi^c \phi^c}{F^2}+\cdots\right) .
\ea
\begin{figure}
\begin{center}
\includegraphics[width=\textwidth]{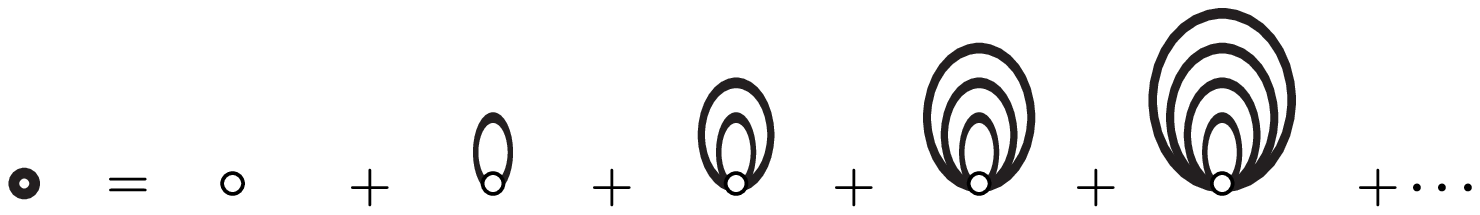}
\end{center}
\caption{The equation for the vacuum expectation
value $V_\phys$. The open thin lined dots indicate an insertion of $-s^0$.
A thick line indicates the full propagator.}
\label{fig:vacuum_gap}
\end{figure}
In this case there is no need for wave function renormalization.
In Fig. \ref{fig:vacuum_gap} we show how the vacuum condensate is given by the
sum of all the tadpole diagrams obtained by contracting the
$\phi^c \phi^c$ fields in all possible ways. 
As explained for the decay constant, the leading in $N$ contribution comes from
the contractions of the same flavour index, i.e. $\phi^c\phi^c$.
Each loop  again implies the integral $A(M^2_\phys)$.
This leads to the following expressions for the vacuum expectation value in 
terms of the low energy constants $F,M^2$ and in terms of the physical
$F_\phys$ and $M^2_\phys$
\be
\label{qbarqlargeN}
V_\phys = V_0 \sqrt{1+\frac{N}{F^2}\overline{A}(M^2_\phys)}\,,
\ee
or
\be
\label{qbarqlargeN2}
V_\phys = \frac{V_0}{\sqrt{1-\frac{N}{F^2_\phys}\overline{A}(M^2_\phys)}}.
\ee

\subsection{$\pi \pi$-scattering amplitude}

The meson-meson scattering amplitude defined in (\ref{defAstu})
in the leading $N$ approximation is somewhat more involved.
There is ample literature on the subject, see for example 
\cite{ColemanlargeN,Dobado:1992jg} and more recently \cite{Kivel1,Kivel3},
which deal with linear or nonlinear massless $O(N+1)/O(N)$ sigma models. 
In the massive case there is the additional complication that tadpoles do not
 vanish. 
As for the previous observables, in order to be leading in $N$, each momentum
 loop
must correspond to a sum over the $N$ flavours. 
We deal with all the generated cactus diagrams in three steps.

First, we consider all insertions on a meson line that do not carry away 
momentum and/or flavour. They can be dealt with simply by using the
full propagator obtained earlier. 

\begin{figure}
\begin{center}
\includegraphics[width=\textwidth]{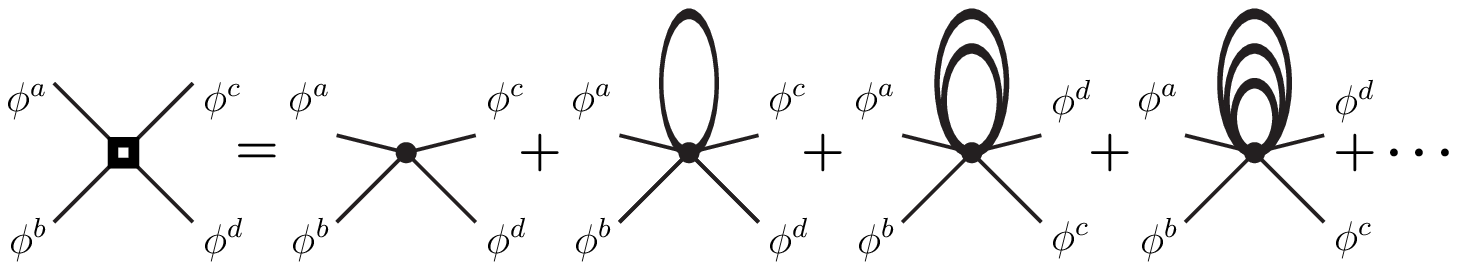}
\end{center}
\caption{\label{fig:pipi1} 
The effective vertex for meson-meson scattering.
The thick lines correspond to the full propagator
produced by the graphs in Fig.~\ref{figgap}.
The large dots are vertices from the Lagrangian (\ref{sigmalag}).}
\end{figure}

Next we deal with effective four-meson couplings, described by the left hand
 side (lhs) of the Eq. in Fig.~\ref{fig:pipi1}. These can be produced by
 resumming all the generalized tadpoles shown on the rhs in 
Fig.~\ref{fig:pipi1}, just
as we did for  $M^2_\phys$, $F_\phys$ and for $V_\phys$, see
 Figs.~\ref{figgap}, \ref{fig:decay_gap} and \ref{fig:vacuum_gap} respectively. 

In the massive non-linear sigma model case, the effective four-$\phi$ couplings
come from both the kinetic and the mass part of the Lagrangian. In the first
representation these may be written as
\be
\mathcal{L}^{(n\ge4)\phi}=
\frac{F^2}{2}\partial_{\mu}\sqrt{1-\frac{\phi^T \phi}{F^2}}
             \partial^{\mu}\sqrt{1-\frac{\phi^T \phi}{F^2}}
 +F^2 M^2\sqrt{1-\frac{\phi^T \phi}{F^2}}-F^2M^2+\frac{M^2}{2}\phi^T\phi.
\ee
Note that we have removed the kinetic terms that give the lowest order
propagator.
The first term may be expanded into 
\be
\mathcal{L}^{(n\ge4)\phi}_{\rm{kin}}=\frac{1}{2F^2}
\frac{\displaystyle   \partial_{\mu}\phi^a \phi^a\partial^{\mu}\phi^b \phi^b}
 {\displaystyle 1-\frac{\phi^T \phi}{F^2}}
=\frac{1}{2F^2}
 \left(\partial_{\mu}\phi^a \phi^a\partial^{\mu}\phi^b\phi^b\right)
 \sum_n{\left( \frac{\phi^T \phi}{F^2}\right)^n}.
\ee
The loops appearing on the rhs of the equation in Fig~\ref{fig:pipi1} 
may be treated in the leading $N$ limit as before. 
The derivatives cannot both appear in the loops at leading order in $N$ 
for the same reasons valid for wave function renormalization. 

The $\mathcal{L}^{(n\ge4)\phi}_{\rm{kin}}$ thus leads to an effective vertex
\be
\label{effectivepipi1}
\frac{1}{2F^2}
\frac{\displaystyle   \partial_{\mu}\phi^a \phi^a\partial^{\mu}\phi^b \phi^b}
 {\displaystyle 1+y}\,
\ee
where $y=\frac{N}{F^2}A(M^2_\phys)$. The loop integral $A(M^2_\phys)$ 
is again produced by the $\phi^T\phi=\phi^c\phi^c$ contractions. 

$\mathcal{L}^{(n\ge4)\phi}_{\rm{mass}}$ may be expanded with
$\sqrt{1-\frac{\phi \phi}{F^2}}
 = \sum^\infty _{n=0}
\left( \begin{array}{c}1/2\\ n \end{array}\right)
\left(-\frac{\phi^T \phi}{F^2} \right)^n$. 
As before, the loops must each come from one $\phi^T\phi$ pair, but now 
we have to take into account the number of ways in which the $\phi^T\phi$ 
pairs can be attached to the four external legs.
For a term  $(\phi^T\phi)^n$ there are $n(n-1)/2$ ways to select four
 outer fields 
and to contract the remaining $n-2$ pairs. If each contraction leads to a
factor $-A(M^2_\phys)$, then for each flavour the $(\phi^T\phi)^n$ term will
contribute $(\phi^T\phi)^2/2\,\,n(n-1) [-A(M^2_\phys)/F^2]^n$. The 
$\mathcal{L}^{(n\ge4)\phi}_{\rm{mass}}$ will then contribute 
$ (\phi^T\phi)^2/2 \sum^\infty _{n=0}
\left( \begin{array}{c}1/2\\ n \end{array}\right) n(n-1)
 [-N A(M^2_\phys)/F^2]^n$,  which is the second derivative
 of $\sqrt{1-NA(M^2_\phys)/F^2}$ with respect to $NA(M^2_\phys)/ F^2$.

\begin{figure}
\begin{center}
\includegraphics[width=\textwidth]{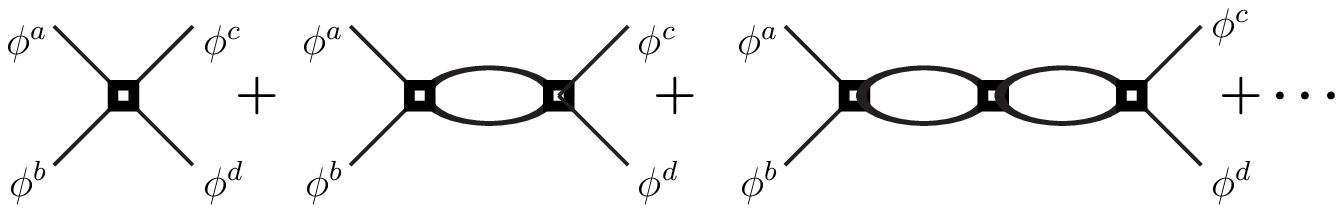}
\end{center}
\caption{\label{fig:pipi2} The remaining diagrams for $A(s,t,u)$.
The vertex is the one obtained from the effective vertex
(\ref{effectivepipi}) with $\delta^{ab}\delta^{cd}$.
}
\end{figure}

The effective four meson vertex coming from the mass term is thus
\be
\label{effectivepipi2}
-\frac{M^2}{8F^2}\frac{(\phi^T\phi)^2}{(1+y)^{(3/2)}}\,.
\ee
Using (\ref{masslargeN}) and (\ref{decaylargeN}) one may write the 
total effective vertex
which has the same form as the lowest order vertex but with physical quantities
\be
\label{effectivepipi}
\frac{1}{2F^2_\phys}\phi^a\partial_\mu\phi^a\,\phi^b\partial^\mu\phi^b
-\frac{M^2_\phys}{8F^2_\phys}\phi^a\phi^a\phi^b\phi^b\,.
\ee
This effective vertex corresponds to the first diagram in Fig.~\ref{fig:pipi2}
 and leads to an amplitude $(p_a+p_b)\cdot(p_c+p_d)-M^2_\phys$. Each
 of these vertices is the building block for the remaining diagrams in
 Fig.~\ref{fig:pipi2}.

We now concentrate on the $A(s,t,u)\delta^{ab}\delta^{cd}$ part of the
 amplitude
defined in (\ref{defAstu}). The corrections to
 $(p_a+p_b)\cdot(p_c+p_d)-M^2_\phys$ 
are generated by the fish-like diagrams given in Fig.~\ref{fig:pipi2}.
This sum is very similar to the sum over bubbles used in the NJL model
\cite{ENJLreview}.
The structure of the vertex is such that for leading $N$ it does not depend
on the loop integral. Each flavour loop in Fig.~\ref{fig:pipi2}
thus adds  a factor
\be
\frac{N(s-M^2_\phys)}{2F^2_\phys} B\left(M^2_\phys,M^2_\phys,s\right)
\ee
with $s=(\mathbf{p}_a+\mathbf{p}_b)^2$ and $B$ the standard two-propagator
 loop integral
\be
B(m^2_,m^2,\mathbf{p}^2)=\frac{1}{i}\int\frac{d^d \mathbf{q}}{(2\pi)^d}
\frac{1}{(\mathbf{q}^2-m^2)((\mathbf{q}-\mathbf{p})^2-m^2)}\,.
\ee
The factor $1/2$ is from the symmetry factor in the loop.
The sum of diagrams forms a geometric series which becomes
\be
A(s,t,u) = \frac{s-M^2_\phys}{F^2_\phys}
\frac{1}{1-\frac{N}{2}\frac{s-M^2_\phys}{F^2_\phys}
               B\left(M^2_\phys,M^2_\phys,s\right)}\,.
\ee
This expression is in agreement\footnote{Compared with \cite{Dobado:1992jg} we
found an extra factor $1/2$ in front of the $B(M^2_\phys,M^2_\phys,s)$
function coming 
from the symmetry factor of the loop. Note that they only worked
to first order in the 
mass.} with both the results in (\ref{resultAstu}) and what was found by
\cite{Kivel1}
in the $M^2\rightarrow 0$ limit. 

\begin{figure}
\begin{center}
\includegraphics[width=0.8\textwidth]{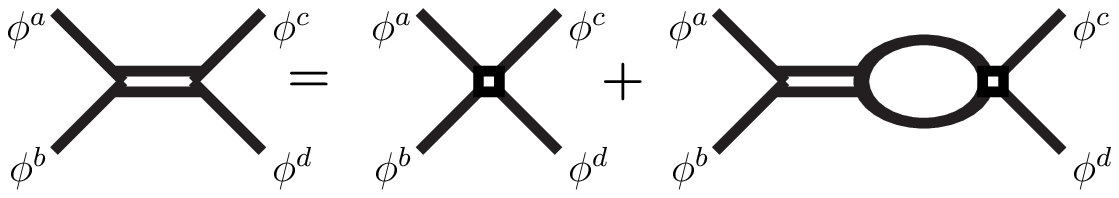}
\end{center}
\caption{The recursive equation which generates all the meson-meson
scattering diagrams. The vertex is the effective vertex of 
(\ref{effectivepipi}). The double line is the full results for $A(s,t,u)$.
The single thick line is the full meson propagator.}
\label{fig:pipi3}
\end{figure}

An alternative way to resum the diagrams in Fig.~\ref{fig:pipi2}
is with a recursion relation, as depicted in Fig.~\ref{fig:pipi3}.
Let's denote $A(s,t,u)$ by a thick double line and a wave function 
renormalized leg by a thick single line (remember that in the large $N$
 limit $Z=1$). The lhs of the equation is then is result we 
sought after. On the rhs we have the renormalized effective $4\phi$ 
vertex plus 
the $A(s,t,u)$ multiplied the renormalized fish diagram. 

As for the mass case, solving the equation by first writing the lowest order 
expression on the rhs, and reinserting the solution into the equation and so on
generates the whole set of diagrams in Fig.~\ref{fig:pipi2}. 

\subsection{Vector and Scalar form factors}

The vector and scalar form factors $F_V$ and $F_S$ in the large $N$ limit
are calculated in much the same manner. One constructs effective vertices
and then sums the diagrams.

\begin{figure}
\begin{center}
\includegraphics[width=\textwidth]{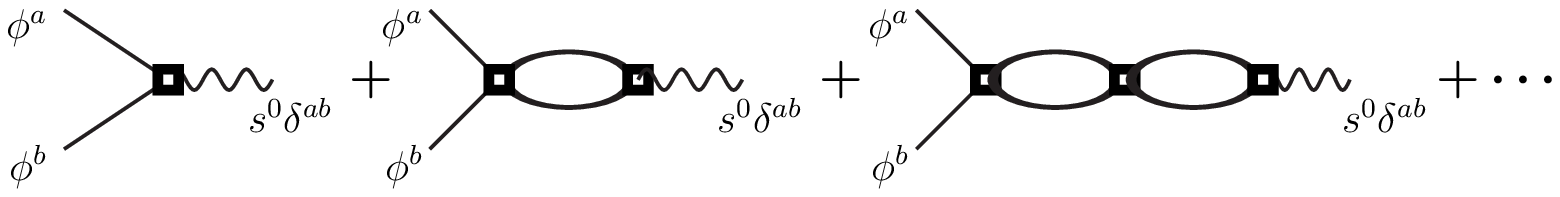}
\end{center}
\caption{\label{fig:Fs} The fish diagrams for the scalar form factor.}
\end{figure}

The result for the vector form factor is particularly simple. 
The vector form factor couplings come from  $\mathcal{L}^{\rm{kin}}$,
 in particular 
from  $F^2(\Phi^Tv_\mu\partial^\mu \Phi-\partial_\mu \Phi^T v^\mu \Phi)$. 
The effective vertex for the first parametrization 
$v^{ab}_\mu\left[ \phi^â \partial^\mu \phi \right]$
is the same as the lowest order vertex and because of the antisymmetry
in flavour indices of $v^{ab}_\mu$ there are also no diagrams
similar to those of Fig.~\ref{fig:Fs}.
The full leading order in $N$ result is thus
\be
F_V(t) = 1\,.
\ee
The scalar current comes from $\mathcal{L}^{\rm{kin}}$, which was discussed
 earlier for the VEV. The sum of all the tadpole diagrams leads to the
 effective vertex 
\be
- B s^0 \frac{\phi^a\phi^a}{\sqrt{1+y}}.
\ee
In the scalar case however, one must also consider fish-like diagrams, see
 Fig.~\ref{fig:Fs}. The arguments used for resumming these diagrams used 
for $\phi \phi$-scattering still apply.

The full result for the scalar form factor is thus
\be
F_S(t) =
\frac{F_S(0)}
{1-\frac{N}{2}
\frac{t-M^2_\phys}{F^2_\phys}B\left(M^2_\phys,M^2_\phys,t\right)}\,,
\ee
with
\be
F_S(0) = -\frac{V_\phys}{F_\phys^2}\,.
\ee
Here we have used the earlier results to write the expression in its simplest
form.

\section{Leading Logarithmic series for $O(N+1)/O(N)$}
\label{LLseries}

In this section we describe the calculation of the leading logarithms and quote
results for various physical quantities.
For a given observable $O_\phys$ we can write the leading logarithm
expansion in many equivalent ways. The two we will use are of the form
\be
\label{defexpL}
O_\phys =  O_0 \left(1+a_1 L +a_2 L^2+\cdots\right)\,,
\ee
with $L$ defined in (\ref{defL}). We will also
expand alternatively in the physical quantities
\be
\label{defexpLphys}
O_\phys =  O_0 \left(1+c_1 L_\phys +c_2 L_\phys^2+\cdots\right)\,,
\ee
with $L_\phys$ defined in (\ref{defLphys}). In both cases we have chosen
the mass scale in the logarithm to be the corresponding mass.

In \cite{paper1} we described how to systematically take into account all
 the necessary 
diagrams for the renormalization of the $\phi$ mass up to five loops.
At each order new diagrams are necessary. To renormalize the mass at one
 loop for instance, 
one must consider the $\mathcal{L}^0$ $4\phi$ vertex and contract two of
 the legs.
As the loop order grows, so does the number of outer legs one must consider and 
the number of possible one loop diagrams contributing.
We give the related discussion here for the decay constant.
The actual calculations were performed by using 
\textsc{FORM} \cite{FORM} extensively.

\subsection{Mass}

The coefficients  $a_i$ of the logs $L^i$ for the mass were calculated in
 \cite{paper1} and are here reproduced for completeness in Tab.~\ref{tabmass1}.
\begin{table}
\begin{center}
\begin{tabular}{|c|c|l|}
\hline
i & $a_i$ for $N=3$ & $a_i$ for general $N$\\
\hline
1 & $-1/2$        & $ 1 - 1/2~N$\\
%\hline
2 & 17/8       & $7/4
          - 7/4~N
          + 5/8~N^2$\\
%\hline
3 & $-103/24$     & $ 37/12
          - 113/24~N
          + 15/4~N^2
          - N^3 $ \\
%\hline
4 & 24367/1152 & $ 839/144
          - 1601/144~N
          + 695/48~N^2
          - 135/16~N^3 $\\
  & & $
          + 231/128~N^4 $ \\
%\hline
5 & $-8821/144$   & $33661/2400
          - 1151407/43200~N
          + 197587/4320~N^2$\\
& & $
          - 12709/300~N^3
          + 6271/320~N^4
          - 7/2~N^5 $ \\
\hline
\end{tabular}
\end{center}
\caption{\label{tabmass1} The coefficients $a_i$ of the leading logarithm $L^i$
up to $i=5$ for the physical meson mass \cite{paper1}.}
\end{table}
The coefficients in terms of fully physical quantities can be derived by using
the results for the decay constant given below.
They are given in Tab.~\ref{tabmass2}.
\begin{table}
\begin{center}
\begin{tabular}{|c|c|l|}
\hline
i & $c_i$ for $N=3$ & $c_i$ for general $N$\\
\hline
%\hline
1 & $ - 1/2$ & $ 1 - 1/2\,N$\\
2 & $ 7/8 $  & $  - 1/4 + 3/4\,N - 1/8\,N^2 $\\
3 & $ 211/48 $ & $  - 5/12 + 7/24\,N + 5/8\,N^2 - 1/16
         \,N^3 $\\
4 & $ 21547/1152 $ & $ 347/144 - 587/144\,N + 47/24\,N^2
          + 25/48\,N^3$\\
 & & $ - 5/128\,N^4 $\\
5 & $ 179341/2304 $ &$  - 6073/1800 + 32351/2400\,N - 
         59933/4320\,N^2$\\
 & & $
       + 224279/43200\,N^3 
  + 761/1920\,N^4 - 7/256\,N^5 $\\
\hline
\end{tabular}
\end{center}
\caption{\label{tabmass2} The coefficients $c_i$ of the leading logarithm
$L^i_\phys$ up to $i=5$ for the physical meson mass.}
\end{table}
The leading logarithm for the masses for $N=3$ at two-loop was first
calculated in \cite{Colangelopipi} and later to full two-loop order
in \cite{Burgi,BCEGS}. Our results agree with those.

In \cite{paper1} we noticed that the expansion of
$M^2/M^2_\phys$ in $L_{M_\phys}$
converged faster than the expansion of $M^2_\phys/M^2$ in terms of $L$.
This was true for both the large $N$ result and the general $N$ case.
{}From the large $N$ result in (\ref{masslargeN2}) we would have naively
expected 
to see a similar improvement in the expansion of $M^2_\phys/M^2$ in terms
of $L_\phys$.
Looking at the coefficients of Tab.~\ref{tabmass2} one can see this is not
the case.
For completeness we also looked at the series of $M^2/M^2_\phys$
in terms of $L_\phys$. The coefficients are of similar size as those
in Tab.~\ref{tabmass2}.

\begin{figure}
\begin{minipage}{0.49\textwidth}
\includegraphics[width=\textwidth]{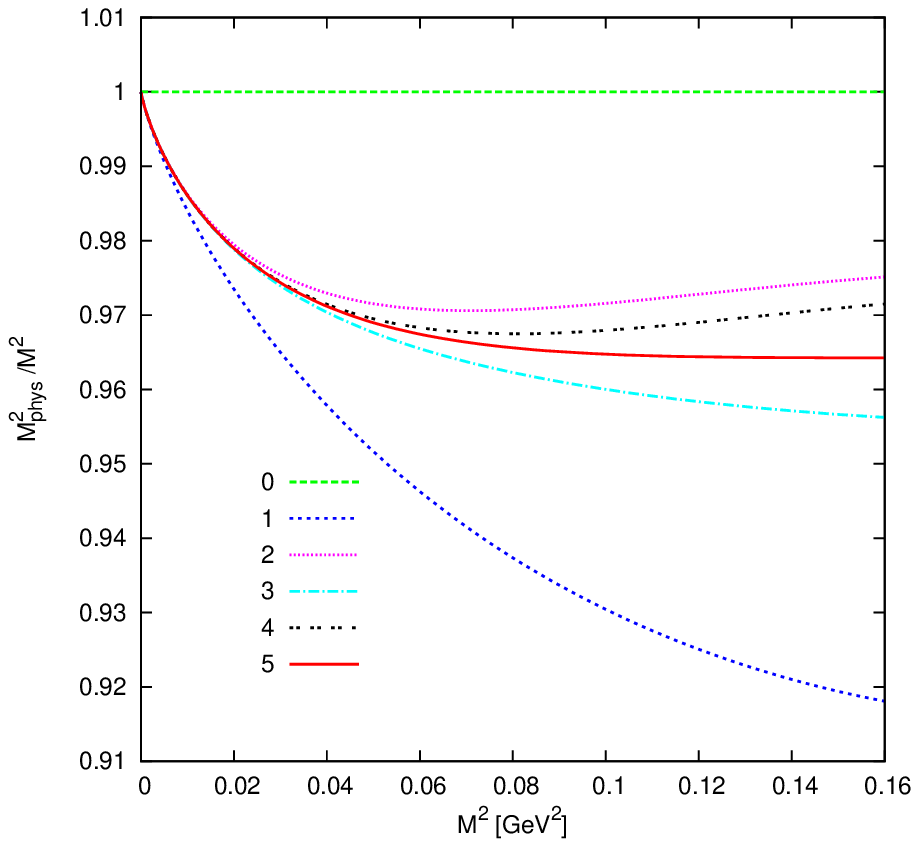}
\centerline{(a)}
\end{minipage}
\begin{minipage}{0.49\textwidth}
\includegraphics[width=\textwidth]{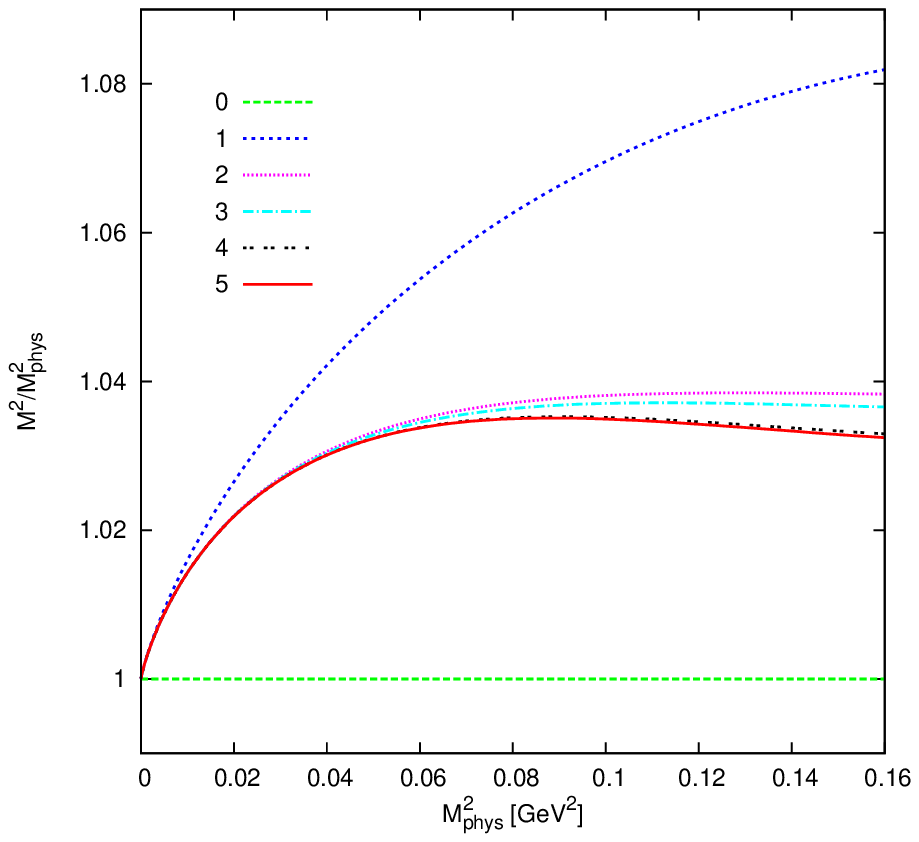}
\centerline{(b)}
\end{minipage}
\caption{\label{figmass1}
The expansions of the leading logarithms order by order for $F=0.090$~GeV,
$\mu =0.77$~GeV and
$N=3$.
(a) $M^2_{\phys}/M^2$, expansion in $L$.
(b) $M^2/M^2_{\phys}$, expansion in $L_{M_\phys}$.}
\end{figure}

We can now use these results to check the expansions and how fast they converge.
In \cite{paper1} the x-axis in Figs. 6(a) and (b) was unfortunately mislabeled.
It should have been $M$~ [GeV] instead of $M^2$~[GeV$^2$].
We have therefore included a similar figure again.
We chose $F=0.090~$GeV and $\mu=0.77~$GeV for the plots presented here
in Fig.~\ref{figmass1}.
\begin{figure}
\begin{minipage}{0.49\textwidth}
\includegraphics[width=\textwidth]{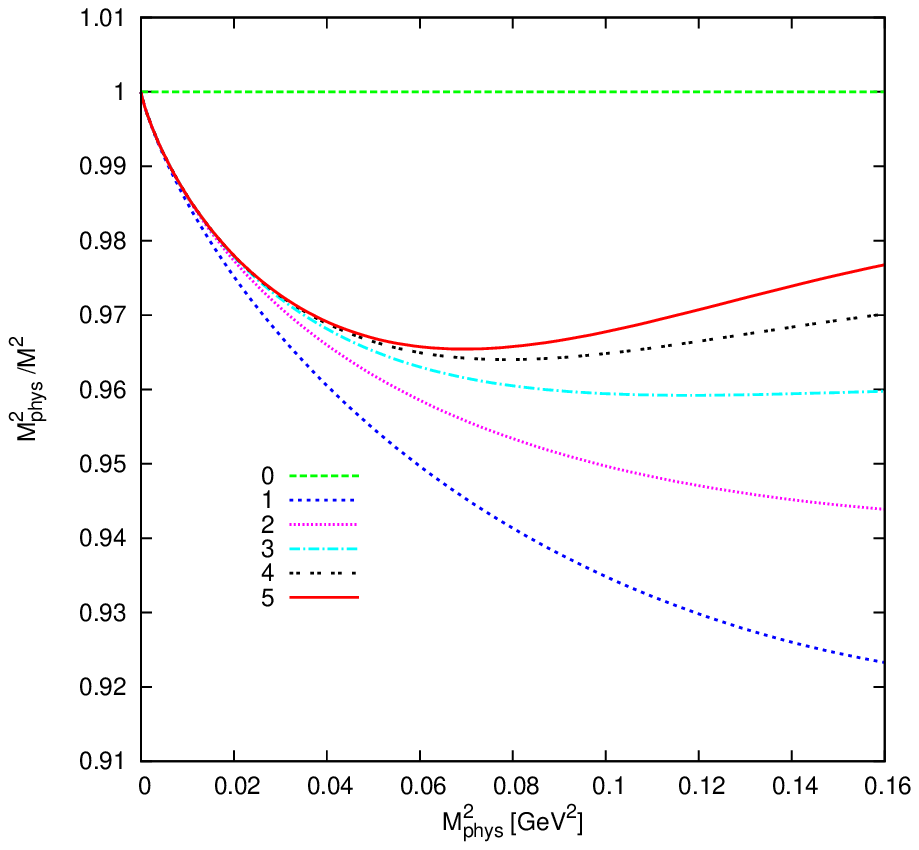}
\centerline{(a)}
\end{minipage}
\begin{minipage}{0.49\textwidth}
\includegraphics[width=\textwidth]{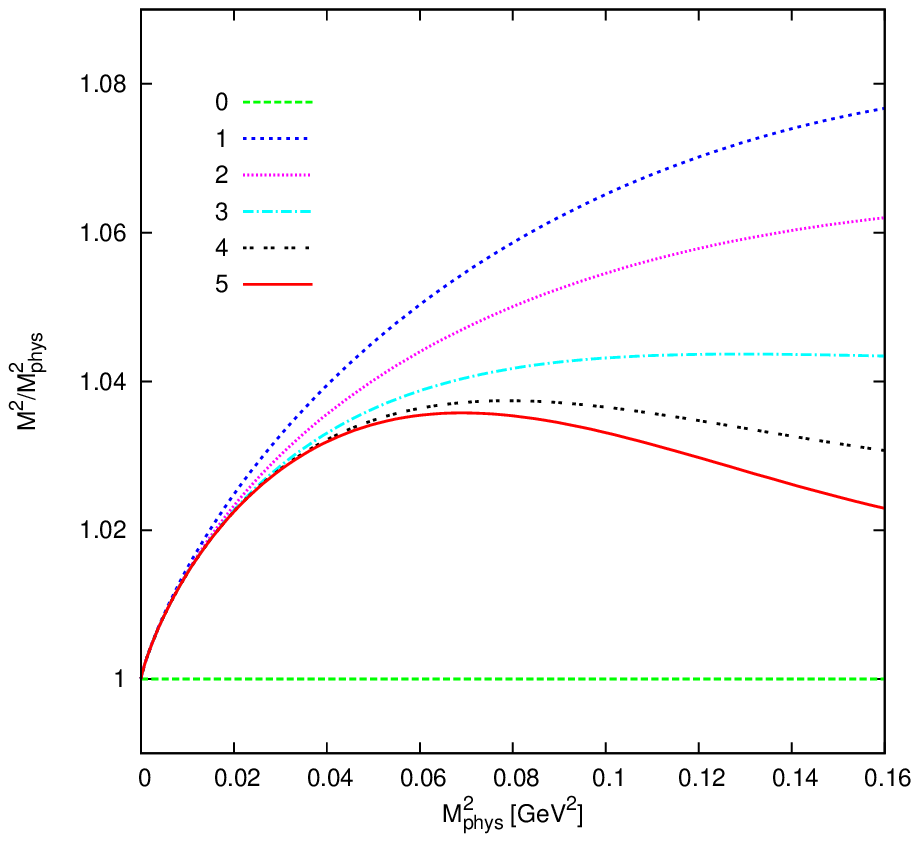}
\centerline{(b)}
\end{minipage}
\caption{\label{figmass2}
The expansions of the leading logarithms order by order for $F_\phys=0.093$~GeV,
$\mu =0.77$~GeV and
$N=3$.
(a) $M^2_{\phys}/M^2$, expansion in $L_\phys$.
(b) $M^2/M^2_{\phys}$, expansion in $L_\phys$.}
\end{figure}
The expansion can also be done in the physical quantities and these
we show as a function of $M^2_\phys$ with $F_\phys$ fixed at $0.093~$GeV
in Fig.~\ref{figmass2}. Both cases have a similar convergence which
is fairly slow for masses above about 300~MeV.

\subsection{Decay constant}
\label{DC}

The decay constant $F_\phys$ is defined in (\ref{defF}). 
We thus need to evaluate a matrix-element with one external axial field 
and one incoming meson.
The diagrams needed for the wave function renormalization were already
evaluated in the calculation for the mass \cite{paper1} since we evaluated the
inverse propagator there.
What remains is thus the evaluation of all relevant 1PI diagrams with an 
external $a^a_\mu$.
\begin{figure}
\begin{center}
\includegraphics[width=\textwidth]{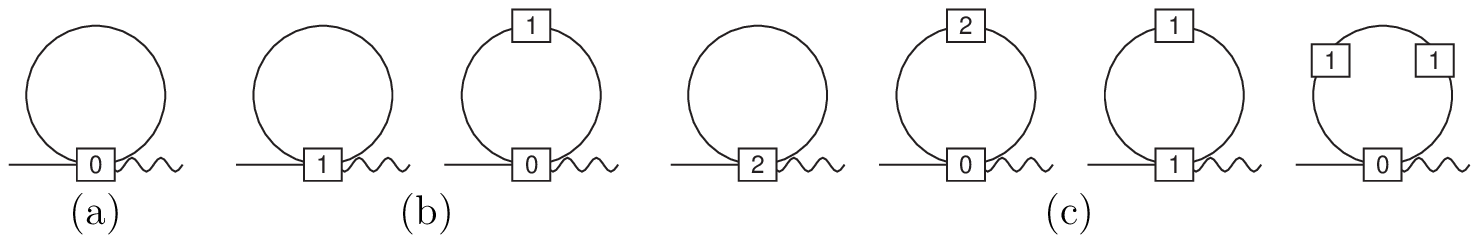}
\end{center}
\caption{\label{fig1PImass} The diagrams needed
up to order $3$ for the one particle irreducible
diagrams with an external meson and an axial current.
The wiggly line indicates the axial vector.
Vertices of order $\hbar^i$ are indicated with \fbox{i}.
(a) The diagram needed at order $\hbar$.
(b) The 2 diagrams needed at order $\hbar^2$.
(c) The 4 diagrams needed at order $\hbar^3$.}
\end{figure}
At order $\hbar$ there is only one diagram, at $\hbar^2$
there are 2 and at order $\hbar^3$ there are 4. These
are shown in Fig.~\ref{fig1PImass}.
We have not shown them but at order $\hbar^4$ there
are 7 and at $\hbar^5$ there are 13 diagrams to be calculated.

\begin{figure}
\begin{center}
\includegraphics[width=\textwidth]{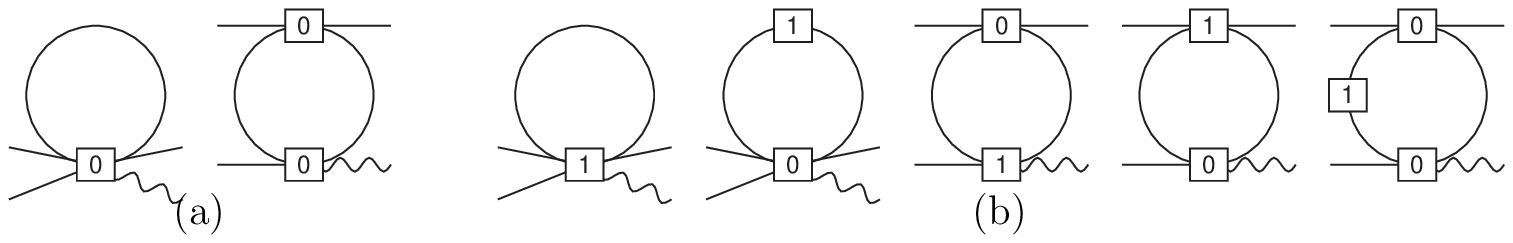}
\end{center}
\caption{\label{fig1PIpipi} The diagrams needed for the divergence
of the 3-meson 1-axial-vector vertex. (a) The 2 diagrams to order $\hbar$.
(b) the 5 diagrams to order $\hbar^2$
}
\end{figure}

To order $\hbar$ it is sufficient to know the lowest-order Lagrangian,
but at order $\hbar^2$ we need to know the (divergent part of the)
vertices coming from the Lagrangian of order $\hbar$ with one and three 
external meson legs and one $a^a_\mu$.
The diagram of Fig.~\ref{fig1PImass}(a) gives the divergence
of the vertex with one meson and one axial vector leg
but we also need to calculate the
divergence of the vertex with three meson legs and one axial vector.
This requires the diagrams shown in Fig.~\ref{fig1PIpipi}(a).

\begin{figure}
\begin{center}
\includegraphics[width=0.5\textwidth]{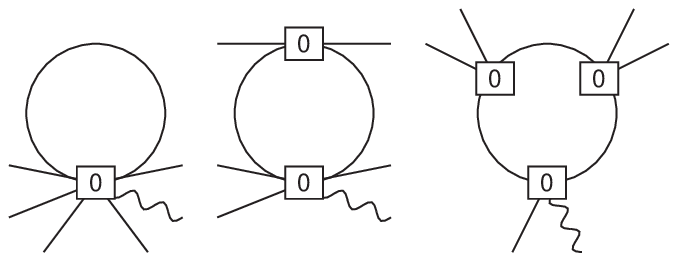}
\end{center}
\caption{\label{fig1PI6pi} The 3 diagrams needed for the divergence
of the 6-meson vector to order $\hbar$.}
\end{figure}

To order $\hbar^3$, we need still more vertices, we need
the divergence of the one meson one axial vector leg vertex
to order $\hbar^2$. These diagrams we have already calculated,
 but we also need the four-leg vertex to order $\hbar^2$ which can be 
calculated from the diagrams in Fig.~\ref{fig1PIpipi}(b). 
Inspection of the vertices there shows we already have all we need but 
for the five meson one axial vector leg vertex at order $\hbar$.
To obtain that we also need to evaluate all diagrams shown in
Fig.~\ref{fig1PI6pi}.
By now, the pattern should be clear,
to get the mass at order $\hbar^n$ in \cite{paper1} we needed
the 2 and four-meson vertex at order $\hbar^{n-1}$,
the 2, 4 and 6-meson vertex at order $\hbar^{n-2}$ and so on.
Here we need for the decay constant to order $\hbar^n$ in addition
the one meson one axial vector and three-meson--one-axial-vector
vertex at order $\hbar^{n-1}$,
the 1, 3 and 5-meson plus one axial vector vertex at order $\hbar^{n-2}$
and so on.
Similarly one can see that to get the decay constant at order $\hbar^n$,
we need to calculate one-loop diagrams with up to $n$ vertices.
The extension  to order $\hbar^5$ shows that we
we need to calculate the 18, 26, 33, 26 and 13 diagrams at orders
 $\hbar^1,\ldots,\hbar^5$ respectively for the mass/wave function 
renormalization and an additional 27, 45, 51, 33 and 13 diagrams
at orders  $\hbar^1,\ldots,\hbar^5$ for the 1PI
diagrams with an axial-vector.

We have organized this calculation by first expanding the
lowest-order Lagrangian to the order needed, up to vertices
with 12 mesons or 11 mesons and one axial vertex.
With these vertices we then calculate all 1PI
diagrams with up to 10 external legs. The divergent part of all
needed integrals can be calculated relatively easily using the
technique described in App.~\ref{integrals} of \cite{paper1}.
At this stage, the dependence on external momenta is also put back as
derivatives on the external legs and everything assembled
to give the divergent part at order $\hbar$ for all the vertices
with up to 10 meson legs or nine mesons plus one axial-vector
using (\ref{hbarnsol}).
So we have assembled everything we need to calculate
the one-loop divergences to order $\hbar^2$. The 26+45 diagrams are
evaluated and we obtain the divergences at order $\hbar^2$
using (\ref{hbarnsol}).
The process is then repeated up to order $\hbar^5$.
All of the above steps have been programmed in FORM.
The CPU time needed increases rapidly with the order $n$ one wishes to reach.
The Lagrangians at higher orders tend to contain very many terms
and constructing the diagrams with many external legs at higher orders
is also extremely time consuming.
The CPU time used on a typical PC for the mass-divergence
to order $\hbar^n$ was approximately 0.1 seconds for $\hbar$, 
0.3 seconds for $\hbar^2$, 11 seconds for $\hbar^3$, 700 seconds for $\hbar^4$
and 30000 seconds for $\hbar^5$ plus a similar amount for the extra diagrams
needed for the decay constant.

We now give the leading logarithms for the decay constant as a function of
$F$ and $M^2$ and of the physical $F_\phys$ and $M_\phys^2$
\ba
F_\phys & = & F  \left(1+a_1 L+ a_2 L^2+\cdots\right)\,,
\nonumber\\
F_\phys & = & F  \left(1+c_1 L_\phys+ c_2 L^2_\phys+\cdots\right)\,.
\ea
The first five $a_i$ coefficients are listed in table \ref{tab:decay}
for the generic $N$ and for the interesting case $N=3$.
The equivalent results for the first five $c_i$ are in table~\ref{tab:decay2}.
Note that once the expression of $F_\phys$ as a function of $F$ is known
one may express the remaining observables as a function of the
physical $M^2_\phys$ and $F_\phys$. This has already been used to calculate the
$c_i$ coefficients in tables ~\ref{tabmass2} and \ref{tab:decay2}
from the corresponding $a_i$.
\begin{table}
\begin{center}
\begin{tabular}{|c|c|l|}
\hline
i & $a_i$ for $N=3$ & $a_i$ for general $N$\\
\hline
%\hline
1 & $ 1$ & $ 
          - 1/2
          + 1/2\,N
          $\\
2 & $ - 5/4$ & $
          - 1/2
          + 7/8\,N
          - 3/8\,N^2
          $\\
3 & $ 83/24 $ & $
          - 7/24
          + 21/16\,N
          - 73/48\,N^2
          + 1/2\,N^3
          $\\
4 & $ - 3013/288 $ & $
           47/576
          + 1345/864\,N
          - 14077/3456\,N^2
          $\\ & & $
          + 625/192\,N^3
          - 105/128\,N^4
          $\\
5 & $  2060147/51840 $ & $
          - 23087/64800
          + 459413/172800\,N
    $\\ & & $
          - 189875/20736\,N^2
          + 546941/43200\,N^3
    $\\ & & $
          - 1169/160\,N^4
          + 3/2\,N^5
          $\\
\hline
\end{tabular}
\end{center}
\caption{\label{tab:decay} The coefficients $a_i$ of the leading logarithm
 $L^i$ for the decay constant $F_\phys$ in the case $N=3$  and
 in the generic $N$ case.}
\end{table}
\begin{table}
\begin{center}
\begin{tabular}{|c|c|l|}
\hline
i & $c_i$ for $N=3$ & $c_i$ for general $N$\\
\hline
%\hline
1 & $ 1 $ & $  - 1/2 + 1/2\,N $\\
2 & $ 5/4 $ & $ 1/2 - 7/8\,N + 3/8\,N^2 $\\
3 & $  13/12 $ & $   - 1/24 + 13/16\,N - 13/12\,N^2 + 5/
         16\,N^3 $\\
4 & $  - 577/288$ & $  - 913/576 + 2155/864\,N - 361/
         3456\,N^2 - 69/64\,N^3$\\
  & & $
         + 35/128\,N^4 $\\
5 & $   - 14137/810 $ & $  535901/129600 - 2279287/172800\,N
          + 273721/20736\,N^2 
    $\\ & & $
     - 11559/3200\,N^3 - 997/1280\,N^4 + 63/256\,N^5 
    $\\
\hline
\end{tabular}
\end{center}
\caption{\label{tab:decay2} The coefficients $c_i$ of the leading logarithm
 $L^i_\phys$ for the decay constant $F_\phys$ in the case $N=3$
and in the generic $N$ case.}
\end{table}

\begin{figure}
\begin{minipage}{0.49\textwidth}
\includegraphics[width=\textwidth]{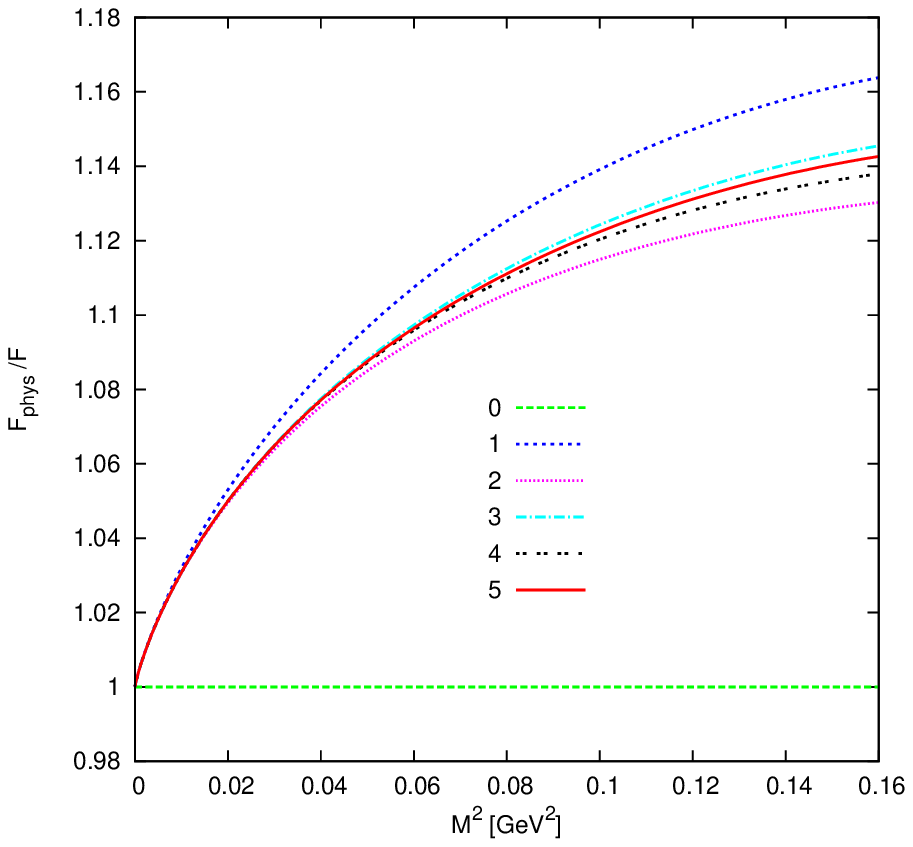}
\centerline{(a)}
\end{minipage}
\begin{minipage}{0.49\textwidth}
\includegraphics[width=\textwidth]{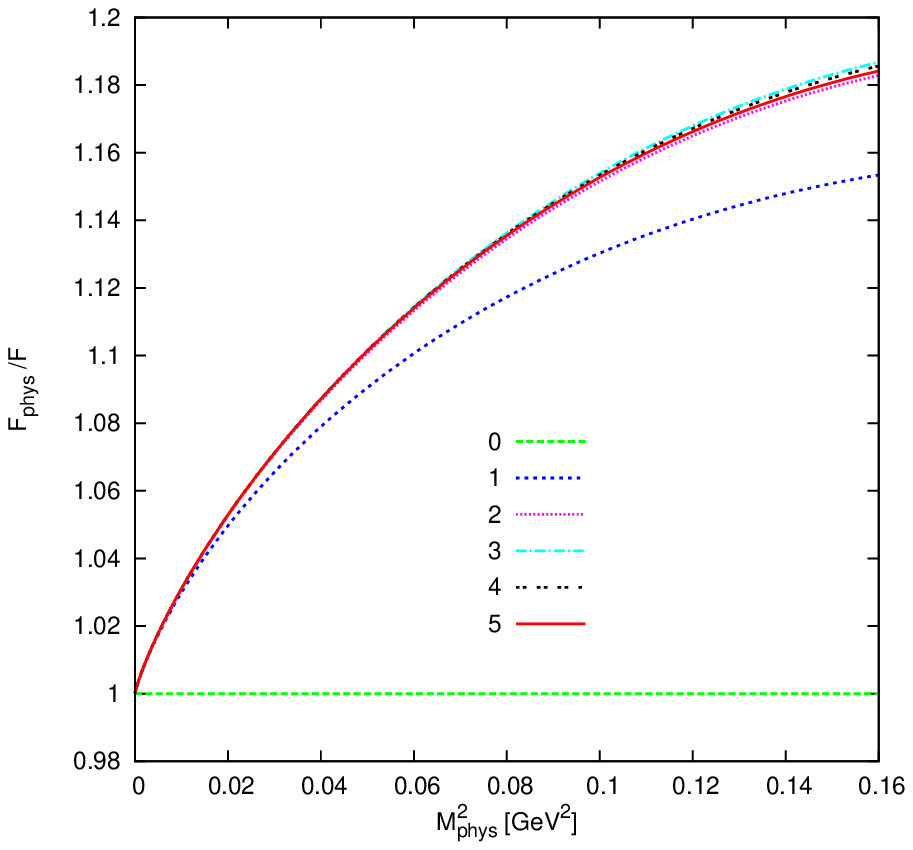}
\centerline{(b)}
\end{minipage}
\caption{\label{figdecay}
The expansions of the leading logarithms order by order for
$\mu =0.77$~GeV and
$N=3$.
(a) $F_\phys/F$ in terms of $M^2$, expansion in $L$ with $F=0.090~$GeV.
(b) $F_\phys/F$ in terms of $M^2_{phys}$, expansion in $L_\phys$
with $F_\phys=0.093~$MeV fixed.}
\end{figure}

We have plotted in Fig.~\ref{figdecay} the expansion in terms of the
unrenormalized quantities and in terms of the physical quantities.
In both cases we get a good convergence but it is excellent for the expansion in
physical quantities.

\subsection{Vacuum expectation value}
\label{vev}

The expression for the leading logarithms of the vacuum expectation
value $V_\phys$ follows from the definition (\ref{defV}).
The diagrams needed and the principle of the calculation can be derived
in the same way as we did for the decay constant in Sect.~\ref{DC}.
The first five coefficients $a_i$ defined by
\be
V_\phys = -2BF^2\left(1+a_1 L+ a_2 L^2+\cdots\right),,
\ee
are given in table \ref{tab:vev} for generic $N$ 
and for the interesting case $N=3$. The $c_i$ for the leading logarithms in
terms of physical quantities are given in table~\ref{tab:vev2}.
\begin{table}
\begin{center}
\begin{tabular}{|c|c|l|}
\hline
n & $a_n$ for $N=3$ & $a_n$ for general $N$\\
\hline
%\hline
1 &     3/2   &  ~$       + 1/2\,N$\\
%\hline
2 &   $- 9/8$  & $       + 3/4\,N
          - 3/8\,N^2$\\
%\hline
3 &   $ 9/2$ &  $       + 3/2\,N
          - 3/2\,N^2
          + 1/2\,N^3 $  \\
%\hline
4 &  $- 1285/128$ &  $        + 145/48\,N
          - 55/12\,N^2
          + 105/32\,N^3
          - 105/128\,N^4 $ \\
%\hline
5 &  $ 46$  &  $      + 3007/480\,N
          - 1471/120\,N^2
          + 557/40\,N^3$\\
 & & $
          - 1191/160\,N^4
          + 3/2\,N^5$\\
\hline
\end{tabular}
\end{center}
\caption{\label{tab:vev} The coefficients $a_i$ of the leading logarithm
 $L^i$ for the VEV $V_\phys$ in the case $N=3$  and in the generic $N$ case.
Note that the coefficients in front of the first subleading $N$ power are often large.}
\end{table}
\begin{table}
\begin{center}
\begin{tabular}{|c|c|l|}
\hline
n & $c_n$ for $N=3$ & $c_n$ for general $N$\\
\hline
%\hline
1 &  3/2     & $1/2\,N$ \\
2 &  21/8    &$
         - 1/4\,N + 3/8\,N^2$ \\
3 & 75/16    &$
        1/4\,N - 1/2\,N^2 + 5/16\,N^3$ \\
4 & 1023/128 &$
        3/16\,N + 5/24\,N^2 - 59/96\,N^3 + 
         35/128\,N^4$ \\
5 & 2669/256 &$
         - 4153/2880\,N + 12299/4320\,N^2
          - 142/135\,N^3$\\
  & & $
          - 167/320\,N^4 + 63/256\,N^5$ \\
\hline
\end{tabular}
\end{center}
\caption{\label{tab:vev2} The coefficients $c_i$ of the leading logarithm
 $L^i_\phys$ for the VEV $V_\phys$ in the case $N=3$  and in the generic
$N$ case.
}
\end{table}

\begin{figure}
\begin{minipage}{0.49\textwidth}
\includegraphics[width=\textwidth]{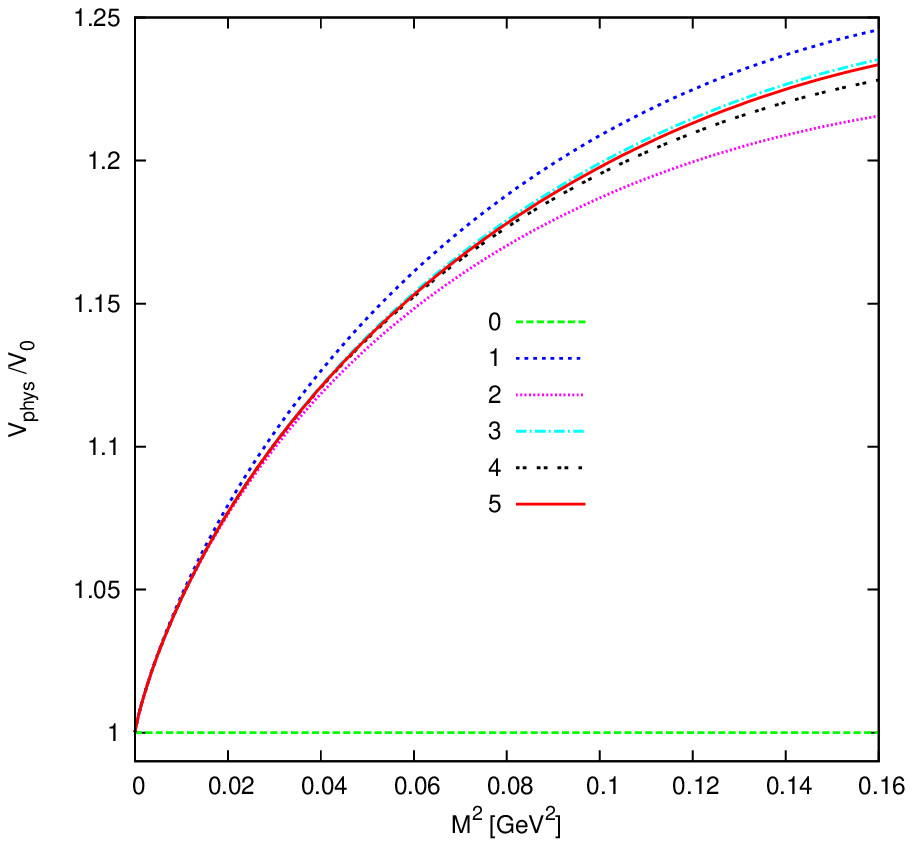}
\centerline{(a)}
\end{minipage}
\begin{minipage}{0.49\textwidth}
\includegraphics[width=\textwidth]{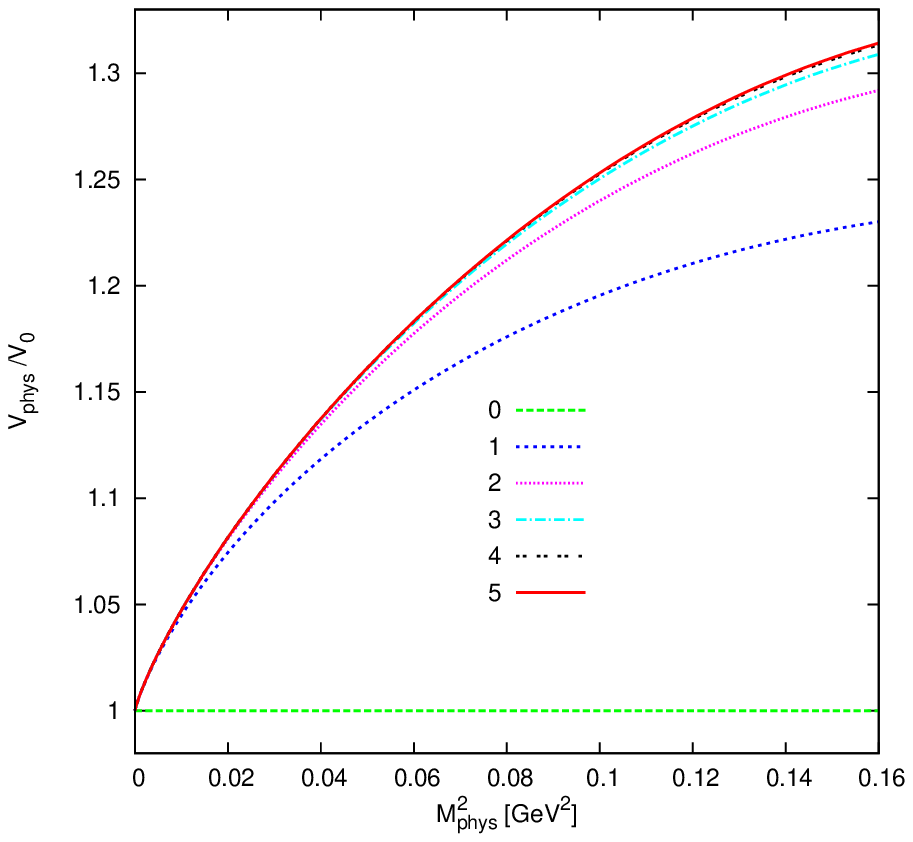}
\centerline{(b)}
\end{minipage}
\caption{\label{figvev}
The expansions of the leading logarithms order by order for
$\mu =0.77$~GeV and
$N=3$.
(a) $V_\phys/V_0$ in terms of $M^2$, expansion in $L$ with $F=0.090~$GeV.
(b) $V_\phys/V_0$ in terms of $M^2_{phys}$, expansion in $L_\phys$
with $F_\phys=0.093~$MeV fixed.}
\end{figure}

We have plotted in Fig.~\ref{figvev} the expansion in terms of the
unrenormalized quantities and in terms of the physical quantities.
In both cases we get a good convergence but it is excellent for the expansion in
physical quantities.

\subsection{$\phi \phi$-scattering amplitude}
\label{pipi}

Elastic $\pi \pi$-scattering is a the ideal interaction to test
the convergence of the ChPT 
expansion since it only involves the $SU(2)_L\times SU(2)_R/SU(2)_V$
Goldstone bosons. It is also the simplest purely strong interaction process.
This interaction has indeed been studied precisely for these reasons first by 
\cite{Weinbergpipi} at tree level and then by \cite{GL1,GL0} at one-loop level. 
One would expect the series to converge rather rapidly,
instead the authors of \cite{GL0} 
found that some one-loop corrections were rather large. Specifically, 
as mentioned in the introduction, for the $a^0_0$ scattering length.
On the other hand, the one-loop LL correction to $a^1_1$ arises only
through $F_{\pi}$ renormalization \cite{Colangelopipi},
which means that the chiral logs are not the main part of the one-loop
correction.
The issue of the convergence of the perturbative ChPT expansion
for $\phi \phi$ scattering is delicate. 

We have above obtained the all-order
leading in large $N$ result. From the calculations for the mass we have already
obtained all the needed divergences to get the LL to four-loop order
for meson scattering. The result can be written as expected in the
form of (\ref{defAstu}). This is the first result where the choice of scale
in the logarithm is not unique. We only quote here the expansion in terms
of physical quantities and define
\ba
\tilde s &=& \frac{s}{M^2_\phys}\,
\nonumber\\
\Delta &=& \frac{t-u}{M^2_\phys}\,
\nonumber\\
L_\mathcal{M} &=&
 \frac{M^2_\phys}{16\pi^2 F^2_\phys}\log\frac{\mu^2}{\mathcal{M}^2}\,.
\ea
with a generic scale $\mathcal{M}$. For scattering lengths an obvious choice
is $\mathcal{M}=M_\phys$ but in the massless case the choice is
$\mathcal{M}^2= s$. Our result for general $N$ is
\ba
\lefteqn{ 
\frac{F^2_\phys}{M^2_\phys}\,A(s,t,u) = 
        \tilde{s}-1+L_\mathcal{M}
  \Big[
 ( 1/6\,\Delta^2 - \tilde{s}^2 + 1/2\,N\,\tilde{s}^2 )
 +( 11/3\,\tilde{s} - N\,\tilde{s} )
}&&\nonumber\\
&& 
 +(  - 8/3 + 1/2\,N ) \Big] 
       + L_\mathcal{M}^2\Big[
 ( 5/96\,\tilde{s}\,\Delta^2 + 181/288\,\tilde{s}^3 + 5/96\,N\,\tilde{s}\,
         \Delta^2
\nonumber\\&&
                  - 163/288\,N\,\tilde{s}^3 + 1/4\,N^2\,\tilde{s}^3 )
       + (  - 5/12\,\Delta^2 - 91/36\,\tilde{s}^2 + 5/12\,N
         \,\Delta^2
\nonumber\\&&
         + 29/18\,N\,\tilde{s}^2 - 3/4\,N^2\,\tilde{s}^2 )
       +  ( 25/18\,\tilde{s} + 23/18\,N\,\tilde{s} + 3/4\,N^2\,
         \tilde{s} )
       +  ( 4/3
\nonumber\\&&
                 - 29/12\,N - 1/4\,N^2 )
\Big]
       + 
L_\mathcal{M}^3 \Big[ ( 361/17280\,\Delta^4 - 317/12960\,\tilde{s}^2\,\Delta^2
\nonumber\\&&
 - 21319/51840\,\tilde{s}^4 - 203/17280\,N\,\Delta^4 + 229/6480\,N\,\tilde{s}^2\,\Delta^2
\nonumber\\&&
 + 28081/
         51840\,N\,\tilde{s}^4 + 1/160\,N^2\,\Delta^4 + 11/1440\,N^2\,\tilde{s}^2\,\Delta^2 - 33/80\,N^2\,
         \tilde{s}^4
\nonumber\\&&
 + 1/8\,N^3\,\tilde{s}^4 )
       + (  - 1901/25920\,\tilde{s}\,\Delta^2 + 51869/
         25920\,\tilde{s}^3 + 3073/25920\,N\,\tilde{s}\,\Delta^2
\nonumber\\&&
 - 49573/25920\,N\,\tilde{s}^3 + 41/288\,N^2\,
         \tilde{s}\,\Delta^2 + 8467/4320\,N^2\,\tilde{s}^3 - 1/2\,N^3\,\tilde{s}^3 )
\nonumber\\&&
       + ( 1283/6480\,\Delta^2 - 907/720\,\tilde{s}^2 - 
         2503/2160\,N\,\Delta^2 - 7193/6480\,N\,\tilde{s}^2
\nonumber\\&&
 + 43/60\,N^2\,\Delta^2 - 3257/1080\,
         N^2\,\tilde{s}^2 + 3/4\,N^3\,\tilde{s}^2 )
       + (  - 1189/1620\,\tilde{s}
\nonumber\\&&
 + 2111/810\,N\,\tilde{s}
          + 607/108\,N^2\,\tilde{s} - 1/2\,N^3\,\tilde{s} )
       +  ( 17/810 + 457/180\,N
\nonumber\\&&
 - 22/5\,N^2 + 1/
         8\,N^3 )
\Big]
       +L_\mathcal{M}^4 \Big[
   ( 1451/1244160\,\tilde{s}\,\Delta^4 + 6457/103680\,\tilde{s}^3\,
         \Delta^2
\nonumber\\&&
 + 61781/248832\,\tilde{s}^5 + 143893/12441600\,N\,\tilde{s}\,\Delta^4 - 77957/
         2073600\,N\,\tilde{s}^3\,\Delta^2
\nonumber\\&&
 - 5387831/12441600\,N\,\tilde{s}^5 - 9089/1382400\,N^2\,\tilde{s}\,
         \Delta^4
 + 5531/230400\,N^2\,\tilde{s}^3\,\Delta^2
\nonumber\\&&
 + 5592583/12441600\,N^2\,\tilde{s}^5 + 1/
         256\,N^3\,\tilde{s}\,\Delta^4 - 1/3840\,N^3\,\tilde{s}^3\,\Delta^2
\nonumber\\&&
 - 5267/21600\,N^3\,\tilde{s}^5 + 1/16
         \,N^4\,\tilde{s}^5 )
       + (  - 6493/77760\,\Delta^4
\nonumber\\&&
 + 9023/103680\,
         \tilde{s}^2\,\Delta^2 - 684899/518400\,\tilde{s}^4 + 43523/345600\,N\,\Delta^4 
\nonumber\\&&
  - 203777/
         1036800\,N\,\tilde{s}^2\,\Delta^2 + 20749/12150\,N\,\tilde{s}^4 - 19091/345600\,N^2\,\Delta^4
\nonumber\\&&
 +          146869/1036800\,N^2\,\tilde{s}^2\,\Delta^2 - 1840297/777600\,N^2\,\tilde{s}^4 + 7/320\,N^3\,
         \Delta^4
\nonumber\\&&
 + 143/5760\,N^3\,\tilde{s}^2\,\Delta^2 + 110897/86400\,N^3\,\tilde{s}^4 - 5/16\,N^4\,
         \tilde{s}^4 )
\nonumber\\&&
       + (  - 680609/1555200\,\tilde{s}\,\Delta^2 + 
         23719/103680\,\tilde{s}^3 - 331117/1555200\,N\,\tilde{s}\,\Delta^2 
\nonumber\\&&
+ 2894959/1555200\,N\,
         \tilde{s}^3 + 16621/86400\,N^2\,\tilde{s}\,\Delta^2 + 2812631/777600\,N^2\,\tilde{s}^3
\nonumber\\&&
 + 77/288\,
         N^3\,\tilde{s}\,\Delta^2 - 153377/86400\,N^3\,\tilde{s}^3 + 5/8\,N^4\,\tilde{s}^3 )
       +  ( 39629/15552\,\Delta^2
\nonumber\\&&
 + 88013/129600
         \,\tilde{s}^2 - 186451/129600\,N\,\Delta^2 - 272671/77760\,N\,\tilde{s}^2
\nonumber\\&&
 - 9227/5400\,N^2\,
         \Delta^2 - 48067/6075\,N^2\,\tilde{s}^2 + 131/120\,N^3\,\Delta^2
\nonumber\\&&
 + 2017/3600\,N^3\,\tilde{s}^2
          - 5/8\,N^4\,\tilde{s}^2 )
       +  ( 667007/48600\,\tilde{s}
\nonumber\\&&
       - 1109347/129600
         \,N\,\tilde{s}
 + 369719/43200\,N^2\,\tilde{s} + 2467/432\,N^3\,\tilde{s}
\nonumber\\&&
 + 5/16\,N^4\,\tilde{s} )
       +  (  - 12349/864
 + 102659/10800\,N + 
         36097/10800\,N^2 
\nonumber\\&&
- 2887/480\,N^3 - 1/16\,N^4 )
\Big]\,.
\label{resultAstu}
\ea
As for all other quantities we see large subleading in $N$ corrections.

From the result (\ref{resultAstu}) we can obtain the different amplitudes
$T^I$ defined in (\ref{defTI}) and project on the partial waves
using (\ref{defTIl}).
The scattering lengths and slopes as defined in (\ref{defscatt})
can then obtained as well and we get 
the LL $L^i_\phys$ in terms of the 
physical $M^2_\phys$ and $F_\phys$.
We have LL contributions for all $a^I_\ell$ up to $\ell=5$
and to the slopes up to $\ell=4$. These we have all
calculated for general $N$.
We give the expansion in 
\be
d^I_{\ell, phys}=d^I_{\ell,tree}(1+c_1 L^1_\phys+c_2 L^2_\phys+\cdots)
\ee 
for the $S$-wave scattering lengths and slopes and 
only quote the phenomenologically relevant case of $N=3$.
As mentioned above a clear choice for the arbitrary scale in the logarithm is
here the physical mass. The lowest order result and the expansion coefficients
are given in table~\ref{tab:scatt_lengths}.
The one- and two-loop results agree with the earlier published ones
\cite{GL0,Colangelopipi,BCEGS}\footnote{The two-loop coefficients
agree with those of \cite{Colangelopipi} except for $b^2_0$. We have checked
that using the full result from \cite{BCEGS} and (3.13) in \cite{Colangelopipi}
reproduces our result.}
\begin{table}
\begin{center}
\begin{tabular}{|c|c|c|c|c|}
\hline
$d^I_\ell$  &     $a^0_0$       &  $a^2_0 $         &   $b^0_0$     &  $b^2_0$\\
\hline
%\hline
\rule{0cm}{0.65cm}$d^I_{\ell,tree} $ & $\frac{7 M^2_\phys}{32\pi F^2_\phys} $
                   & $\frac{-2 M^2_\phys}{32\pi F^2_\phys} $
                   & $\frac{8 M^2_\phys}{32\pi F^2_\phys} $
                   & $\frac{-4 M^2_\phys}{32\pi F^2_\phys} $ \\
%\hline
\rule{0cm}{0.45cm}$c_1$ &         9/2         &  $ - 3/2 $           &      26/3       & $ - 10/3$  \\
%\hline
$c_2$ &    857/42         &  $ - 31/6    $ &     1871/36     & $ - 169/36$ \\
%\hline
$c_3$ &   153211/1512 &  $ - 7103/216$  &   2822/9      & $ - 352/9 $ \\
%\hline
$c_4$ &    41581/84      & $ - 7802/45 $  &  744463/43  & $ - 1309703/6480$ \\
\hline
\end{tabular}
\end{center}
\caption{\label{tab:scatt_lengths} The coefficients $c_i$ of the
leading logarithm series
$d^I_{\ell, phys}=d^I_{\ell,tree}(1+c_1L^1_\phys+c_2L^2_\phys+\cdots)$ for the 
$\mathcal{A}_{\pi \pi \rightarrow \pi \pi}$ for the
 scattering lengths, $d^I_\ell=a^0_{0},a^2_{0}$,  and for the slopes
 $d^I_\ell=b^0_0,b^2_0$ in the case $N=3$. All in units of $M_\phys$.}
\end{table}

\begin{figure}
\begin{minipage}{0.49\textwidth}
\includegraphics[width=\textwidth]{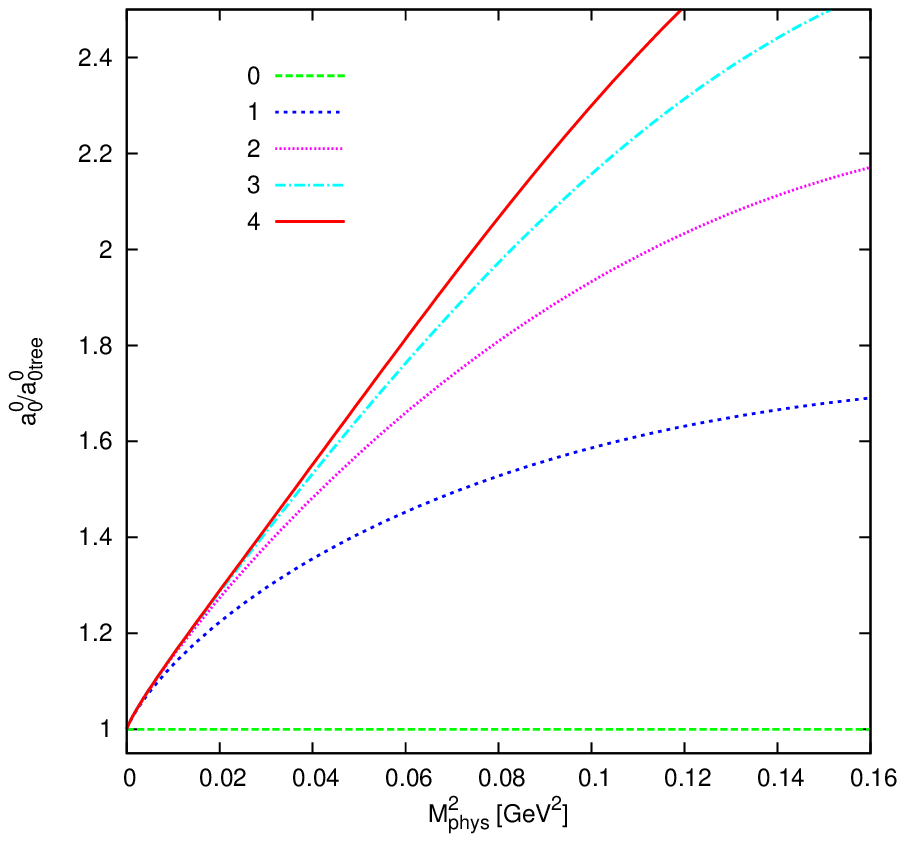}
\centerline{(a)}
\end{minipage}
\begin{minipage}{0.49\textwidth}
\includegraphics[width=\textwidth]{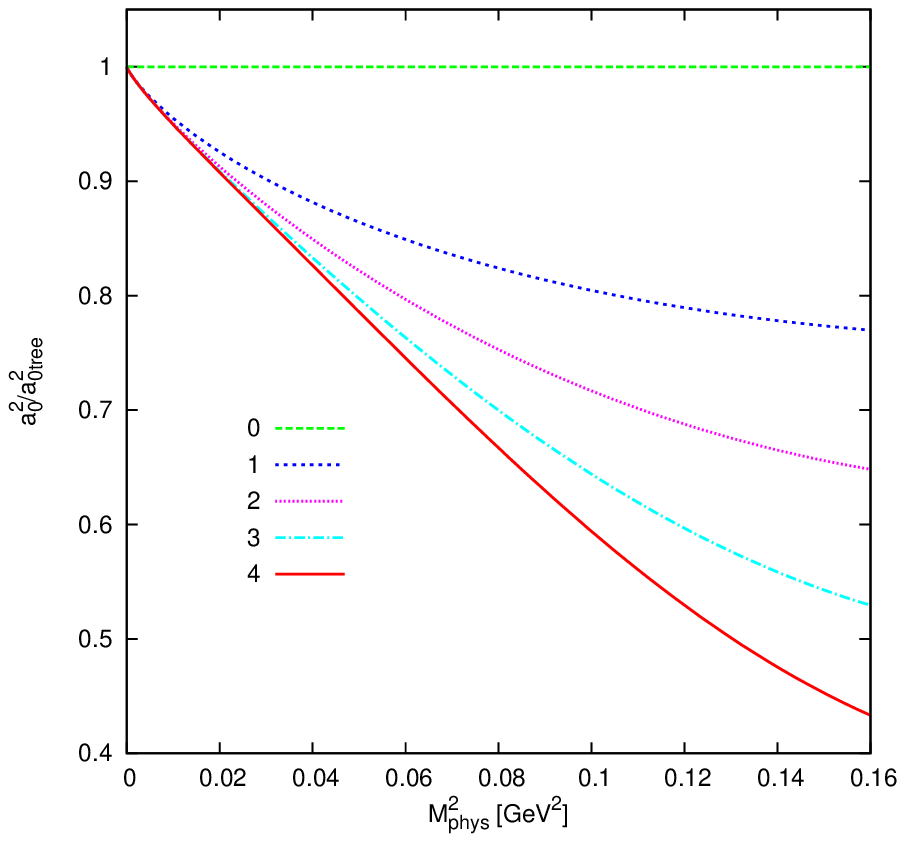}
\centerline{(b)}
\end{minipage}
\caption{\label{figaij}
The expansions of the leading logarithms order by order for
$\mu =0.77$~GeV and
$N=3$.
(a) $a^0_0/a^0_{0\mathrm{tree}}$
(b)  $a^2_0/a^2_{0\mathrm{tree}}$ in terms of $M^2_{phys}$,
expansion in $L_\phys$
with $F_\phys=0.093~$MeV fixed.}
\end{figure}

We have plotted in Fig.~\ref{figaij} the expansion in terms of the
physical quantities of $a^0_0$ and $a^2_0$. There is an excellent convergence
for mass up to $0.2$~GeV but above $0.3$~GeV it becomes rather slow for $a^0_0$.
For $a^2_0$ it is somewhat better but also rather slow at the higher masses.

\begin{figure}
\begin{minipage}{0.49\textwidth}
\includegraphics[width=\textwidth]{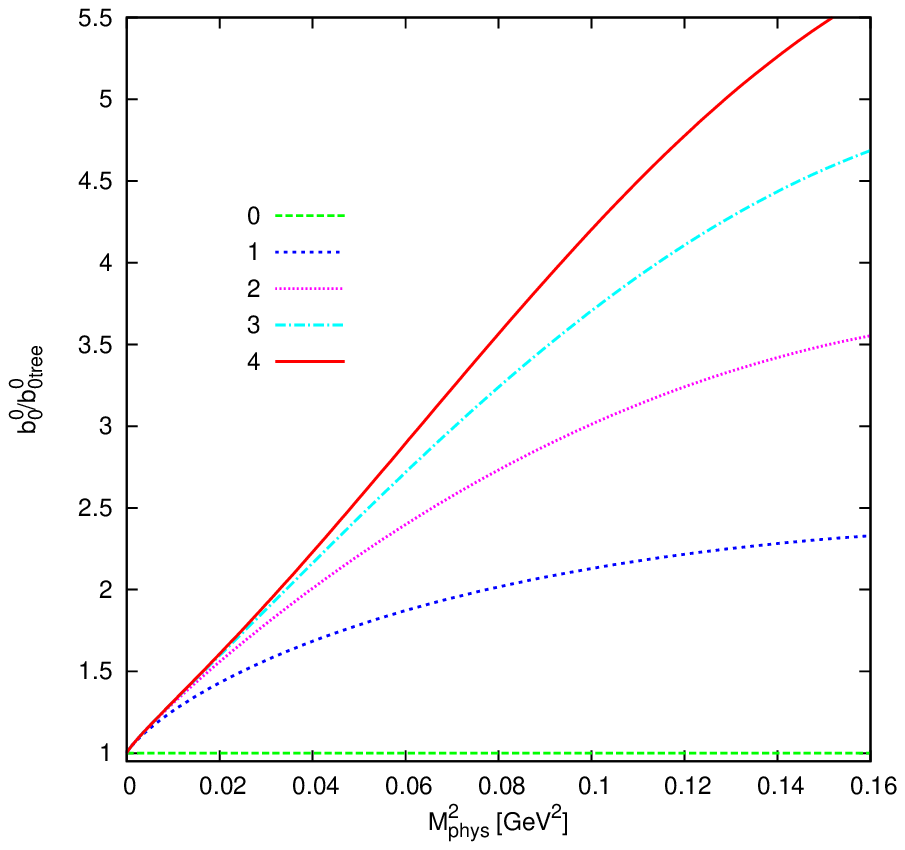}
\centerline{(a)}
\end{minipage}
\begin{minipage}{0.49\textwidth}
\includegraphics[width=\textwidth]{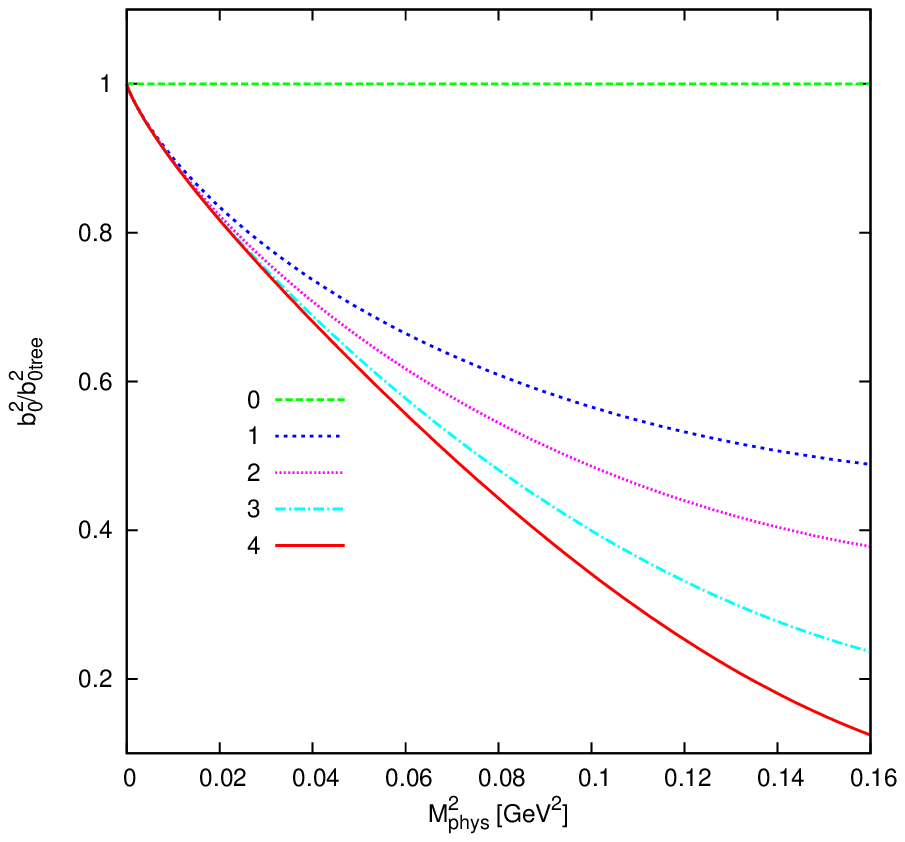}
\centerline{(b)}
\end{minipage}
\caption{\label{figbij}
The expansions of the leading logarithms order by order for
$\mu =0.77$~GeV and
$N=3$.
(a) $b^0_0/b^0_{0\mathrm{tree}}$
(b)  $b^2_0/b^2_{0\mathrm{tree}}$ in terms of $M^2_{phys}$,
expansion in $L_\phys$
with $F_\phys=0.093~$MeV fixed.}
\end{figure}

In Fig.~\ref{figbij} we plotted the expansion in terms of the
physical quantities of the slopes $b^0_0$ and $b^2_0$.
There is an excellent convergence
for mass up to $0.25$~GeV but above the convergence slower for $b^0_0$. 
$b^2_0$ converges better but also rather slow at the high mass
end. 

In the massless case we can obtain the coefficients at higher orders also
with different methods \cite{Kivel1,Kivel3} and our result agrees with those.
We can also use our result in this limit to test the often used
elastic unitarity arguments. The partial waves $T^I_\ell$ satisfy under the
assumption of elastic unitarity
\be
\label{elastic}
\Im\left[T^I_\ell\right] = \frac{\sqrt{s}}{2q} \left|T^I_\ell\right|^2\,.
\ee
In the massless case the scale of the logarithm should be related to $s$
and we know that this should come in the combination
$l(\mu^2/s)= \ln|\mu^2/s|+i\pi\theta(s)$.
The leading logarithm part can be written as
\be
\label{expandTIl}
T^I_\ell = \sum_{n=0,\infty} e_n^I s^{n+1} l(\mu^2/s)^n\,.
\ee
In the chiral limit the interaction must vanish at $s=0$.
Inserting (\ref{expandTIl}) in (\ref{elastic}) and keeping only the
leading logarithm part allows to determine all the higher coefficients
in terms of the lowest order one.
\be
\label{unitarity1}
e_n^I = \pi\left(\frac{e_0^I}{\pi}\right)^{n+1}\,.
\ee
Note that in the massless case there is really no regime where elastic unitarity
(\ref{elastic}) is valid. We are simply testing here how much
of the leading logarithm in this case follows from the so-called righthand
two-body cut. The result (\ref{unitarity1}) can be written explicitly
\be
\label{unitarity2}
T^I_\ell = \frac{e^I_0 s}{1-\frac{e^I_0 s}{\pi}l(\mu^2/s)}\,.
\ee
In table \ref{tab:unitarity} we compare for $N=3$ the exact coefficients derived
from (\ref{resultAstu}) with those from (\ref{unitarity2}) for the expansion
\be
\label{expandTIl2}
T^I_\ell(s) = e^I_0 s\left(1+ f_1 \frac{s}{16\pi^2F^2_\phys}l(\mu^2/s)
+ f_2\left(\frac{s}{16\pi^2F^2_\phys}l(\mu^2/s)\right)^2+\cdots\right)\,.
\ee
It is clear from the table that the assumption of elastic unitarity does
not give a good approximation to the LL in the chiral limit.
\begin{table}
\begin{center}
\begin{tabular}{|c|cc|cc|cc|}
\hline
  &\multicolumn{2}{c|}{ $T^0_0$} 
  &\multicolumn{2}{c|}{ $T^1_1$} 
  &\multicolumn{2}{c|}{ $T^2_0$}\\
\hline
\rule{0cm}{0.5cm}
$e^I_\ell$ & \multicolumn{2}{c|}{$\frac{1}{16\pi F_\phys^2}$}
           & \multicolumn{2}{c|}{$\frac{1}{32\pi F_\phys^2}$}
           & \multicolumn{2}{c|}{$\frac{-1}{32\pi F_\phys^2}$}\\[2mm]
\hline
  & exact &\hspace*{-6mm}elastic & exact &\hspace*{-5mm}elastic   & exact & \hspace*{-5mm}elastic \\
\hline
$f_1$ &25/18         & 1 & 0             & 1/2 & $- 10/9$      & $-1/2$\\
$f_2$ &25/18         & 1 & 5/4           & 1/4 & $35/36$       & $ 1/4$\\
$f_3$ &18637/9720    & 1 & $- 901/3240$  & 1/8 &$- 50077/48600$& $-1/8$\\
$f_4$ &540707/291600 & 1 & 207871/170100 & 1/16& 134077/145800 & $ 1/16$\\
\hline
\end{tabular}
\end{center}
\caption{\label{tab:unitarity} The coefficients $e^I_\ell$ and $f_i$
defined in (\ref{expandTIl2}) for the exact LL in the chiral limit
and those derived using the assumption of elastic unitarity for
the case $N=3$ and $T^0_0$, $T^1_1$ and $T^2_0$.}
\end{table}

\subsection{Form factors $F_S$ and $F_V$}
\label{FF}

The vector and scalar form factors were defined in \ref{defFS} and \ref{defFV}. 
The procedure to find the LL for this observable follow the same
lines of the one for the decay constant, with the difference
that in representation 1 the vertex
between the meson fields and the vector current is simply given by
\be
\mathcal{L}_{\mathrm{int}}= (1/2)\,v^{ab}_{\mu}\left[\partial^\mu \phi_a   \phi_b 
             - \phi_a \partial_\mu \phi_b \right].
\ee
For the wave function renormalization one may again use the results
obtained for the mass calculation.
As in the previous subsection we express here the results
in terms of $\tilde t= t/M^2_\phys$ and $L_\mathcal{M}$ with a scale
$\mathcal{M}^2$ some combination of $t$ and $M^2_\phys$.
The result for $F_V$ to four-loop-order for the LL is:
\ba
\label{resultFV}
\lefteqn{F_V(t) = 1+ L_\mathcal{M} \Big[
           1/6\,\tilde t
          \Big]
       +  L_\mathcal{M}^2  \Big[
          \tilde t\,(- 11/12
          + 5/12\,N)
          +\tilde t^2\,( 5/36
          - 1/24\,N)
           \Big]}&&\nonumber\\&&
       +  L_\mathcal{M}^3   \Big[
         \tilde t \,( + 1387/648
          - 845/324\,N
          + 7/9\,N^2)
         + \tilde t^2\, (- 4007/6480
\nonumber\\&&
          + 3521/6480\,N
          - 29/180\,N^2)
         + \tilde t^3\, (+ 721/12960
          - 47/1440\,N
          + 1/80\,N^2)
          \Big]
\nonumber\\&&
       + L_\mathcal{M}^4   \Big[
          \tilde t \,(- 44249/15552
          + 222085/31104\,N
          - 55063/10368\,N^2
\nonumber\\&&
          + 127/96\,N^3)
         + \tilde t^2 \,(+ 349403/155520
          - 15139/4860\,N
          + 86719/51840\,N^2
\nonumber\\&&
          - 199/480\,N^3)
         + \tilde t^3 \,(- 85141/155520
          + 885319/1555200\,N
\nonumber\\&&
          - 5303/19200\,N^2
          + 21/320\,N^3)
         +\tilde t^4\, ( + 4429/103680
          - 57451/1555200\,N
\nonumber\\&&
          + 289/14400\,N^2
          - 1/240\,N^3)
           \Big]\,.
\ea
Note that $F_V(0) =1$ as it should be.
We can extract from this the expansion for the radius and curvature
defined in (\ref{defradii}).
These are given in table~\ref{tab:radii} in terms of the expansion
in $L_\phys$ for the physical case $N=3$. The general coefficients can be
easily derived from (\ref{resultFV}). The dash indicates that this
cannot appear to a given order for the LL.
\begin{table}
\begin{center}
\begin{tabular}{|c|c|c||c|c|}
\hline
n & $ \langle r^2 \rangle_V$ & $c_V$&$\langle r^2\rangle_S$&$c_S$\\
\hline
%\hline
$c_1$  & 1          & ---            & $6$          & --- \\
%\hline
$c_2$  & 2          & 1/72           & $- 29/3 $    & 43/36 \\
%\hline
$c_3$  & 853/108    & $-71/162$      & $- 581/54$   & $- 727/324$  \\
%\hline
$c_4$  & 50513/1296 & $- 25169/7776$ & $- 75301/648$& 4369/810  \\
\hline 
\end{tabular}
\end{center}
\caption{\label{tab:radii} 
The coefficients $c_i$ of the expansion in $L_\phys$
in the expansion of the radii $\langle r^2 \rangle_{V,S}$ and 
the curvature $c_{V,S}$  in the $N=3$ case.}
\end{table}
The result up to two-loop order agrees with the LL extracted from the full
two-loop calculation \cite{BCT}. We do not present numerical results
for the vector form factor since these are dominated in the physical
case $N=3$ by the large higher order coefficient contributions,
see e.g. \cite{GL1,BCT}.

For the scalar form factor $F_S(t)$ defined in (\ref{defFS}) we have already
done the calculations we need to four-loop-order during the calculation for
the VEV $V_\phys$ to five-loop-order. We write the result in the form
\be
F_S(t) = F_S(0) f_S(t)\,.
\ee
The value for $F_S(0)$ can be obtained from the calculation or via
the Feynman-Hellman theorem in (\ref{FeynmanHellmann}).
The latter allows for the $a_i$ coefficients for $F_S(0)$ expanded
in terms of $L$ to be derived easily from table \ref{tabmass1}.
We have checked that both methods agree.
In table \ref{tab:FS0} we quote the coefficients $c_i$
of
\be
F_S(0) = 2B\left(1+c_1L+c_2 L^2+\cdots\right)\,,
\ee
for the case $N=3$ and general $N$.
\begin{table}
\begin{center}
\begin{tabular}{|c|c|l|}
\hline
 & $c_i$ for $N=3$ & $c_i$ for general $N$\\
\hline
$c_1 $ & $ - 1 $ & $ 2
          - N $\\
$c_2 $ & $ 31/8 $ & $
           5/4
          - 1/4\,N
          + 3/8\,N^2
          $\\
$c_3$ & $ 65/6 $ & $
          - 5/3
          + 41/12\,N
          + N^2
          - 1/4\,N^3
          $\\
$c_4$ & $ 76307/1152 $ & $
           655/144
          - 901/144\,N
          + 341/48\,N^2
          + 17/48\,N^3
$\\&&$
          + 11/128\,N^4
          $\\
\hline
\end{tabular}
\end{center}
\caption{\label{tab:FS0} The coefficients $c_i$ of the leading logarithm
 $L^i_\phys$ for the scalar form factor at zero momentum transfer $F_S(0)$
 in the case $N=3$  and in the generic
$N$ case.}
\end{table}

We can now express $f_S(t)$ using the same notation as for $F_V(t)$.
\ba
\label{resultFS}
\lefteqn{
f_S(t) = 1 +L_\mathcal{M} \Big[
          \tilde t \,(- 1/2
          + 1/2\,N)
          \Big]
       +L_\mathcal{M}^2 \Big[
           \tilde t \,( 1/18
          + 7/36\,N
          - 1/4\,N^2)
}&&\nonumber\\&&
           +\tilde t^2 \,( 11/72
          - 29/72\,N
          + 1/4\,N^2)
          \Big]
       + L_\mathcal{M}^3 \Big[
          \tilde t \,( 599/648
          - 181/324\,N
\nonumber\\&&
          - 53/108\,N^2
          + 1/8\,N^3)
          + \tilde t^2 \,( 275/1296
          - 427/648\,N
          + 301/432\,N^2
\nonumber\\&&
          - 1/4\,N^3)
          + \tilde t^3 \,(- 89/864
          + 7/24\,N
          - 271/864\,N^2
          + 1/8\,N^3)
          \Big]
\nonumber\\&&
       +L_\mathcal{M}^4 \Big[
           \tilde t \,(- 10981/3888
          + 37373/7776\,N
          - 3733/2592\,N^2
          - 23/48\,N^3
\nonumber\\&&
          - 1/16\,N^4)
           +\tilde t^2 \,(- 22859/28800
          + 89951/48600\,N
          - 823067/777600\,N^2
\nonumber\\&&
          - 4021/21600\,N^3
          + 3/16\,N^4)
           +\tilde t^3 \,(- 959/32400
          + 82529/259200\,N
\nonumber\\&&
          - 1421/2025\,N^2
          + 51877/86400\,N^3
          - 3/16\,N^4)
          + \tilde t^4 \,( 76459/1555200
\nonumber\\&&
          - 70997/388800\,N
          + 423961/1555200\,N^2
          - 121/600\,N^3
          + 1/16\,N^4)
          \Big]\,.
\nonumber\\&&
\ea
{}From (\ref{resultFS}) we can derive the expression for the scalar radius and
curvature defined in (\ref{defradii}). The expansion coefficients $c_i$
in terms of the physical logarithm $L_\phys$ are given in table~\ref{tab:radii}
for the physical case $N=3$. The general case can be easily obtained from
(\ref{resultFS}). The coefficients of the LL extracted from the full
two-loop calculation of \cite{BCT} agree.

\begin{figure}
\begin{minipage}{0.49\textwidth}
\includegraphics[width=\textwidth]{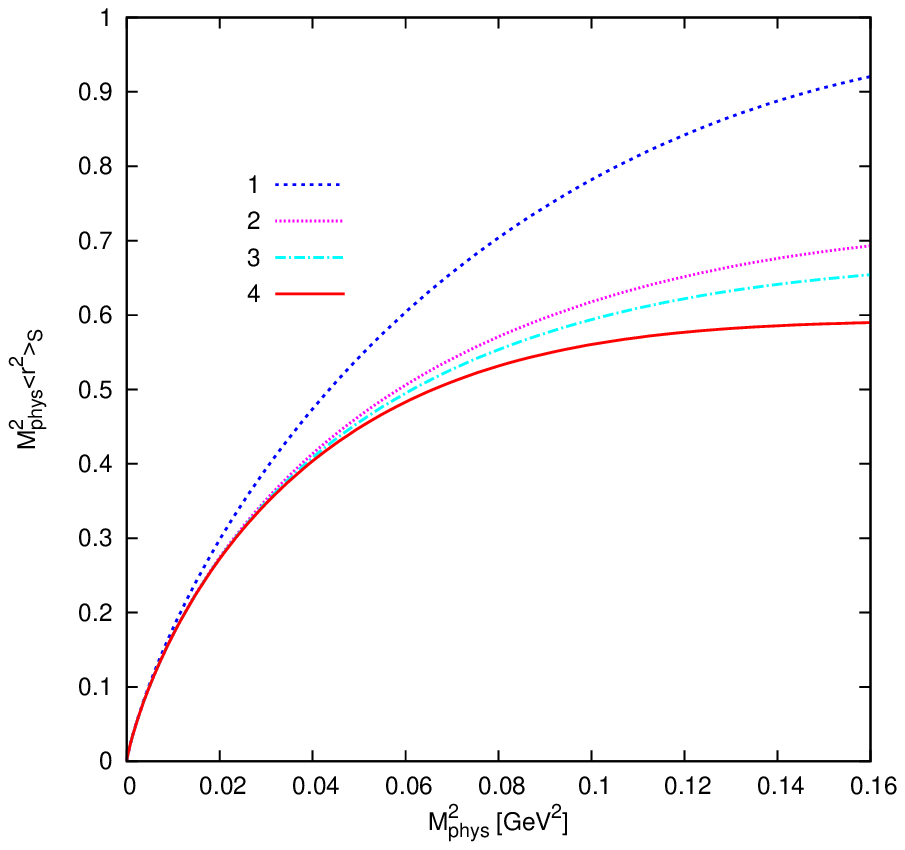}
\centerline{(a)}
\end{minipage}
\begin{minipage}{0.49\textwidth}
\includegraphics[width=\textwidth]{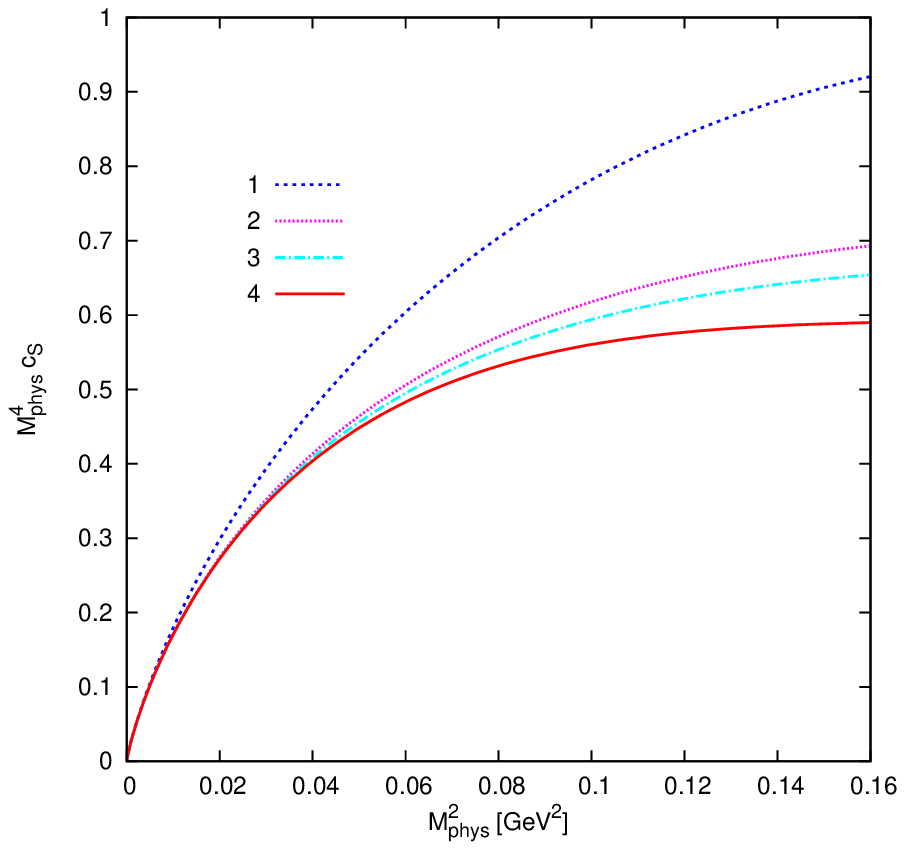}
\centerline{(b)}
\end{minipage}
\caption{\label{figFS}
The expansions of the leading logarithms order by order for
$\mu =0.77$~GeV and
$N=3$.
(a) $M^2_\phys\langle r^2\rangle_S$
(b)  $M^4_\phys c_S$ in terms of $M^2_{phys}$,
expansion in $L_\phys$
with $F_\phys=0.093~$MeV fixed.}
\end{figure}

In Fig.~\ref{figFS} we plotted the expansion in terms of the
physical quantities of the radius and curvature.
There is an excellent convergence
for masses up to $0.3$~GeV but above the convergence is slower
in both cases.

\section{Conclusions}

In this work we extended our previous work on the massive nonlinear
$O(N+1)/O(N)$ sigma model to many more observables. 
We calculated the leading logarithmic corrections to the
decay constant and the vacuum expectation value to five-loop-order
and to the scalar and 
vector form factors and meson-meson scattering to four-loop order
for generic $N$. We used these results to extract scattering lengths and slopes
and have presented the physically most relevant cases for $N=3$ of these.
Results for all other cases have been obtained but would have added
significantly to the length of the paper.

Our original hope had been to find a pattern behind the coefficients of the LL
and to make an all order conjecture for the leading LL.
We succeeded in deriving such an expression for the leading term in the
large $N$ limit but we found no general expression.

The large $N$ approximation, as we already noted in \cite{paper1},
is a surprisingly poor approximation of the  LL series for all of the
observables we considered. For example, looking at the five-loop results,
the first neglected term, the $N^4$ term, often has a large coefficient
compared with the $N^5$ term. For this term to be 
negligible, i.e. a 10\% correction of the leading term, $N$ must be large,
in some cases $N>20$.  This is understandable if one considers that
the subleading $N^{i<n}$ 
terms in the coefficients come from non-cactus diagrams and different flavour
routings of the cactus diagrams. 
Though each of these diagrams is suppressed by $1/N$ in the large $N$ limit,
the number of diagrams and the number of ways to route the flavour
structure seem to produce large coefficients for the subleading in $N$
terms.

We have also performed some numerical test of the convergence with parameter
values of the range needed for two-flavour ChPT. For masses around $0.1~$GeV the
convergence for all quantities studied is excellent. It is reasonable
for most quantities up to about $0.3$~GeV as can be seen on the various plots
but becomes unstable around $0.4$~GeV for some of the quantities studied.

\section*{Acknowledgments}

This work is supported by the 
Marie Curie Early Stage Training program “HEP-EST” (contract number
MEST-CT-2005-019626),
European Commission RTN network,
Contract MRTN-CT-2006-035482  (FLAVIAnet), 
European Community-Research Infrastructure
Integrating Activity
``Study of Strongly Interacting Matter'' (HadronPhysics2, Grant Agreement
n. 227431)
and the Swedish Research Council.

\appendix

\section{Integrals}
\label{integrals}



%% References

%% References with bib Tex database:

%\bibliographystyle{model1a-num-names}

\end{document}